\documentclass{iopart}
\usepackage{graphicx}
\usepackage{bm}
\expandafter\let\csname equation*\endcsname\relax
\expandafter\let\csname endequation*\endcsname\relax
\usepackage{amsmath}
\usepackage{amssymb}
\usepackage[dvips, colorlinks, pdfborder={0 0 0}, plainpages=false]{hyperref}

\newcommand{\affiliation}[1]{\address{#1}}

\newcommand{\sub}[1]{_{\rm #1}}

\newcommand{\ddt}[1]{\frac{d{#1}}{dt}}
\newcommand{\dddt}[1]{\frac{d^2{#1}}{dt^2}}
\newcommand{\dtn}[2]{\frac{d^{#2}{#1}}{dt^{#2}}}

\newcommand{\tbar}{\bar{t}}
\newcommand{\xbar}{\bar{x}}

\newcommand{\half}{\frac{1}{2}}
\newcommand{\xten}[1]{\times 10^{#1}}

\newcommand{\map}[1]{\ensuremath{\mathcal{M}_{\rm #1}}}
\newcommand{\distorted}[1]{\hat{#1}}
\newcommand{\inertial}[1]{\bar{#1}}
\newcommand{\newgrid}[1]{\breve{#1}}
\newcommand{\newdistorted}[1]{\grave{#1}}
\newcommand{\grid}[1]{#1}

\newcommand{\frameB}[1]{\tilde{#1}}     

\newcommand{\Caltech}{\affiliation{$^2$Theoretical Astrophysics 103-33,
    California Institute of Technology, Pasadena, CA 91125}}
\newcommand{\Cornell}{\affiliation{$^1$Center for Radiophysics and Space
    Research, Cornell University, Ithaca, New York, 14853}}
\newcommand{\Fullerton}{\affiliation{$^3$Gravitational Wave Physics and Astronomy Center, California State University Fullerton, Fullerton, CA 92831}}

\begin{document}

\title[Dynamical Excision Boundaries in Spectral Evolutions of BBH Spacetimes]{
Dynamical Excision Boundaries in Spectral Evolutions of Binary Black Hole Spacetimes
}


\author{Daniel A. Hemberger$^1$, Mark A. Scheel$^2$, Lawrence E. Kidder$^1$, B\'ela Szil\'agyi$^2$,
Geoffrey Lovelace$^3$, Nicholas W. Taylor$^2$, and Saul A. Teukolsky$^1$}\Cornell\Caltech\Fullerton

\date{\today}

\begin{abstract}
  Simulations of binary black hole systems using the Spectral
  Einstein Code (SpEC) are done on a computational domain that excises
  the regions inside the black holes.  It is imperative that the
  excision boundaries are outflow boundaries with respect to the
  hyperbolic evolution equations used in the simulation.
  We employ a time-dependent mapping between the fixed computational
  frame and the inertial frame through which the black holes move.
  The time-dependent parameters of the mapping are adjusted throughout
  the simulation by a feedback control system in order to follow the
  motion of the black holes, to adjust the shape and size of the
  excision surfaces so that they remain outflow boundaries, and to
  prevent large distortions of the grid.  We describe in detail the
  mappings and control systems that we use. We show how these
  techniques have been essential in the evolution of binary black hole
  systems with extreme configurations, such as large spin magnitudes
  and high mass ratios, especially during the merger, when apparent
  horizons are highly distorted and the computational domain becomes
  compressed.  The techniques introduced here may be useful in other
  applications of partial differential equations that involve
  time-dependent mappings.
\end{abstract}


\maketitle

\section{Introduction}

Feedback control systems are ubiquitous in technological
applications. They are found, for example, in thermostats, autopilots,
chemical plants, and cruise control in automobiles.  The purpose of a
control system is to keep some measured output (such as the
temperature in a room) at some desired value by adjusting some input
(such as the power to a furnace).

In the last few years, feedback control systems have also found
applications in the field of numerical relativity, particularly in
simulations of binary black hole systems that employ spectral methods
and excision
techniques~\cite{Scheel2006,Boyle2007,Scheel2009,Szilagyi:2009qz,Buchman:2012dw}.

Black hole excision is a means of avoiding the physical singularities
that lurk inside black holes. The idea is to solve Einstein's
equations only in the region outside apparent horizons, cutting out
the region inside the horizons.  The boundaries of the excised regions
are called excision boundaries.  Causality ensures that the excision
boundaries and the excised interiors cannot affect the physics of the
exterior solution, and an appropriate hyperbolic formulation of
Einstein's equations~\cite{Lindblom2004,Lindblom2006} can ensure that
gauge and constraint-violating degrees of freedom also do not
propagate out of the excised region.

Excision is straightforward for black holes that remain stationary in
the coordinates that are used in the simulation, but excision becomes
more complicated when the black holes move or change shape.  For
numerical methods based on finite differencing, the excision
boundaries can be changed at every time step by activating or
deactivating appropriate grid points and adjusting differencing
stencils~\cite{Szilagyi:2004gi,Pretorius2005a,Motamed:2006uw,Szilagyi2007, bbhprl98a, Brandt2000, Alcubierre2001, Shoemaker2003}.
However, for spectral
numerical methods, there is no equivalent of deactivating individual
grid points; instead, spectral methods are defined in finite extended
spatial regions with smooth boundaries.  The nearest equivalent to the
finite-difference excision approach would be interpolating all
variables to a new slightly offset grid at every time step, which
would be computationally expensive.
Therefore, spectral numerical
methods need a different approach to reconcile the need for moving
black holes with the need for a fixed excision boundary inside of each
black hole.

The solution~\cite{Scheel2006} to this problem adopted by our group
makes use of multiple coordinate systems.  We call ``inertial
coordinates'' those coordinates that asymptotically correspond to an
inertial observer; in these coordinates the black holes orbit each
other, have distorted shapes, and approach each other as energy is
lost to gravitational radiation.  Spectral methods are applied in
another coordinate system, ``grid coordinates'', in which the excision
boundaries are spherical and stationary.  We connect grid coordinates
with inertial coordinates by means of an analytic mapping function
$\cal M$ that depends on some set of time-dependent parameters
$\lambda(t)$.  These parameters must be continually adjusted so that each
spherical, stationary grid-frame excision boundary is mapped to a
surface in the inertial frame that follows the motion and the shape of
the corresponding black hole as the system evolves.  It is this
adjustment of each parameter $\lambda(t)$ 
that is accomplished by means of feedback control systems,
one control system per parameter.

In this paper, we describe in detail the mapping functions and the
corresponding feedback control systems that we use to handle black
hole excision with spectral methods.  Some earlier implementations
have been described previously~\cite{Buchman:2012dw, Scheel2006, Boyle2007, Szilagyi:2009qz, Chu2009},
but there have
been many improvements that now allow spectral excision methods to
produce robust simulations of binary black hole systems, including
those with unequal masses, high spins, and precession.  We describe
these improvements here.  In Sec.~\ref{sec:ControlTheory}
we review control theory
and present simple examples of control systems.
In Sec.~\ref{sec:Implementation}
we discuss the implementation of control systems in the SpEC~\cite{SpECwebsite}
code.  Sections~\ref{sec:control-systems-maps}
and~\ref{sec:size-control}
detail
the coordinate mappings used in SpEC and
the feedback control systems used to control them.
In Sec.~\ref{sec:merger-ringdown} we describe the transition to the
post-merger domain with a single excision boundary.
In Sec.~\ref{sec:non-map-systems}
we describe applications of control systems in SpEC besides the ones used
to adjust map parameters.  We summarize in Sec.~\ref{sec:summary}.

\section{Control Theory}
\label{sec:ControlTheory}

To motivate control theory, we begin with a simple example: cruise control
in an automobile.  Suppose we wish to control the speed $v$ of a car
that is driving up an incline of angle $\theta$.  The equation of
motion for this system is
\begin{equation}
 \frac{dv}{dt} = -\frac{\eta}{m} v + \frac{F}{m} - g\sin\theta,
 \label{eq:CruiseControlEOM}
\end{equation}
where $m$ is the mass of the car, $g$ is the gravitational acceleration,
$\eta$ is a drag coefficient, and $F$ is a force supplied by the car's engine.
We wish to determine $F$ so as to cause the car to maintain a speed of
$v=v_0$, even if the angle $\theta$ changes as the car climbs the incline.
To do this, we choose the force at time $t$ to be:
\begin{equation}
 \frac{F(t)}{m} = K_P Q(t) + K_I \int_0^t Q(\tau)\, d\tau,
 \label{eq:CruiseControlSignal}
\end{equation}
where the control system is turned on at time $t=0$, where
$K_I$ and $K_P$ are constants, and
where $Q(t) = v_0 - v(t)$
is the quantity that we wish to drive to zero. We call
$Q$ the control error. Substituting Eq.~\eqref{eq:CruiseControlSignal} 
into Eq.~(\ref{eq:CruiseControlEOM}) and differentiating with respect to
time yields
\begin{equation}
\frac{d^2 Q}{dt^2} + \left(K_P+\frac{\eta}{m}\right) \frac{dQ}{dt} 
+ K_I Q = g \cos\theta\frac{d\theta}{dt},
\end{equation}
which is the equation for a damped, forced harmonic oscillator. 
By choosing $K_P$ to produce critical damping, and choosing $K_I$ to
set a timescale, this choice will drive $v$ toward $v_0$ as desired.

\begin{figure}[h!]
  \begin{center}
      \includegraphics[height=2.0cm]{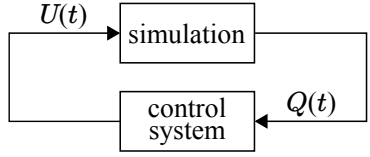}
  \end{center}
  \caption{A generic control circuit.
  The simulation outputs a measure of error, $Q$, which is used by
  the control system. The control system then outputs a signal, $U$,
  which changes the behavior of the simulation.
  }
  \label{fig:circuit}
\end{figure}

The basic structure of a control system is easily understood as a
feedback loop (see Fig.~\ref{fig:circuit}).  The simulation (e.g., a binary
black hole simulation, or a car driving up an incline) produces some
measure of error, $Q(t)$, that defines the deviation from some desired
target value.  This error acts as the input for the control
system, which then computes a control signal $U(t)$ (e.g., the
derivative of a map parameter, or the force supplied by a car's
engine) that will minimize the error $Q(t)$ when fed back into the
simulation.

A simple and effective way to compute $U(t)$ is to make it a linear
combination of the error, $Q(t)$, and integrals and/or derivatives of
the error.  The term proportional to the error acts to reduce the
deviation from the desired value, the terms proportional to
derivatives of the error act to oppose rapid deviations, and the terms
proportional to the integrals act to reduce any persistent deviation or
offset that accumulates over time.  In the cruise control example,
we used a proportional and an integral term only in
Eq.~(\ref{eq:CruiseControlSignal}).

We now turn to another example of a control system that is more closely
related to the way we use control systems in binary black hole simulations.
Consider two coordinate systems, $(x,t)$ and $(\xbar,\tbar)$, that are
related by the map
\begin{align}
	x &= \xbar - V(t)\,\tbar, \label{eq:ExampleMap} \\
	t &= \tbar.
\end{align}
We wish to control the parameter $V(t)$ so that a wave 
$f(\xbar,\tbar)=f(\xbar-c\tbar)$ that propagates at speed $\bar{v}=c$ in the
$(\xbar,\tbar)$ coordinates will propagate at some arbitrary desired
speed $v_d$ in the $(x,t)$ coordinates.  According to Eq.~\eqref{eq:ExampleMap},
the speed of the wave in the $(x,t)$ coordinates is $v=c-V(t)$.  Therefore
we define a control error $Q$ to be
\begin{equation}
Q = c-V(t)-v_d,
\end{equation}
and we construct a control system that drives this control error to zero.
If we choose the control signal $U(t)$ to be $d^2 V/d t^2$, then the simplest
feedback loop that can be constructed uses only a term proportional to
the error and amplified by a ``gain'' $K_P$.
Then the evolution of $V(t)$ is given by
\begin{equation}
\dddt{V} = K_PQ = K_P\left[c-V(t)-v_d\right].
\end{equation}
The solution to this equation is of the form
\begin{equation}
V(t) = c-v_d + A_1\sin(\alpha t) + A_2\cos(\alpha t),
\end{equation}
where $\alpha := \sqrt{K_P}$, which is oscillatory for $K_P > 0$ and
divergent for $K_P < 0$.  Thus we see that adding only a proportional
term to the control signal $U(t)$ is insufficient, 
since it does not reduce the control error $Q$.

However, if we add a derivative term to the feedback equation,
\begin{equation}
\dddt{V} = K_PQ + K_D\ddt{Q},
\end{equation}
then the solution is of the form
\begin{equation}
V(t) = c-v_d + e^{-\half K_D t}\left[B_1\sin(\beta t) 
+ B_2\cos(\beta t)\right],
\end{equation}
where $\beta := \half\sqrt{4K_P-K_D^2}$.
This solution is stable with an exponentially damped envelope when
$4K_P \geq K_D^2$, which will cause $v\rightarrow v_d$ as 
$t\rightarrow\infty$.

\begin{figure}[h!]
  \begin{center}
      \includegraphics[height=5cm]{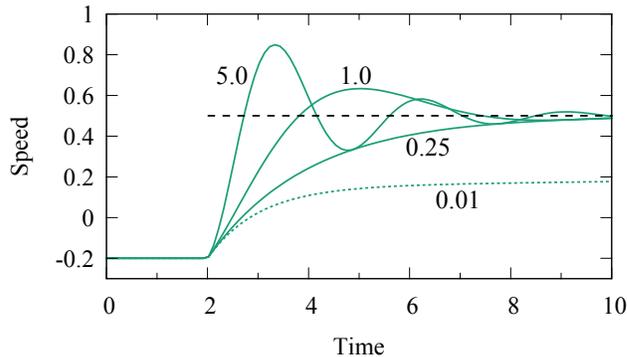}
  \end{center}
  \caption{The speed in $(x,t)$ coordinates, $v$, is plotted for a family of
  gains $K_P$ with $K_D=1$ fixed. The wave velocity is $c=-0.2$ and the
  desired velocity is $v_d=0.5$ (dashed line). The controller turns on at
  $t=2$. For gains in the stable region, where $K_P \geq 0.25$, $v$ settles
  down to $v_d$. One overdamped solution
  with $K_P = 0.01$ is plotted for comparison (dotted line).}
  \label{fig:simple}
\end{figure}

Notice that this control system allows us to choose a $V(t)$ such that
$v$ is the opposite sign of $c$ and the wave is left-going in the
$(x,t)$ frame instead of right-going. This behavior can be seen in the
example in Fig.~\ref{fig:simple}.

The overdamped solution (dotted line in Fig.~\ref{fig:simple}) has a 
persistent offset that can be ameliorated by adding an integral term to
the feedback equation, as was done in the cruise control example.
In principle we could continue to add more terms,
but in practice, it is usually sufficient to use a PID
(proportional-integral-derivative) controller, which has terms proportional
to the control error, its integral, and its derivative.
When the underlying system is unknown, this is the best controller to use 
\cite{Bennett93}.

\section{Control Systems in SpEC}
\label{sec:Implementation}

\begin{figure}[h!]
  \begin{center}
      \includegraphics[height=4cm]{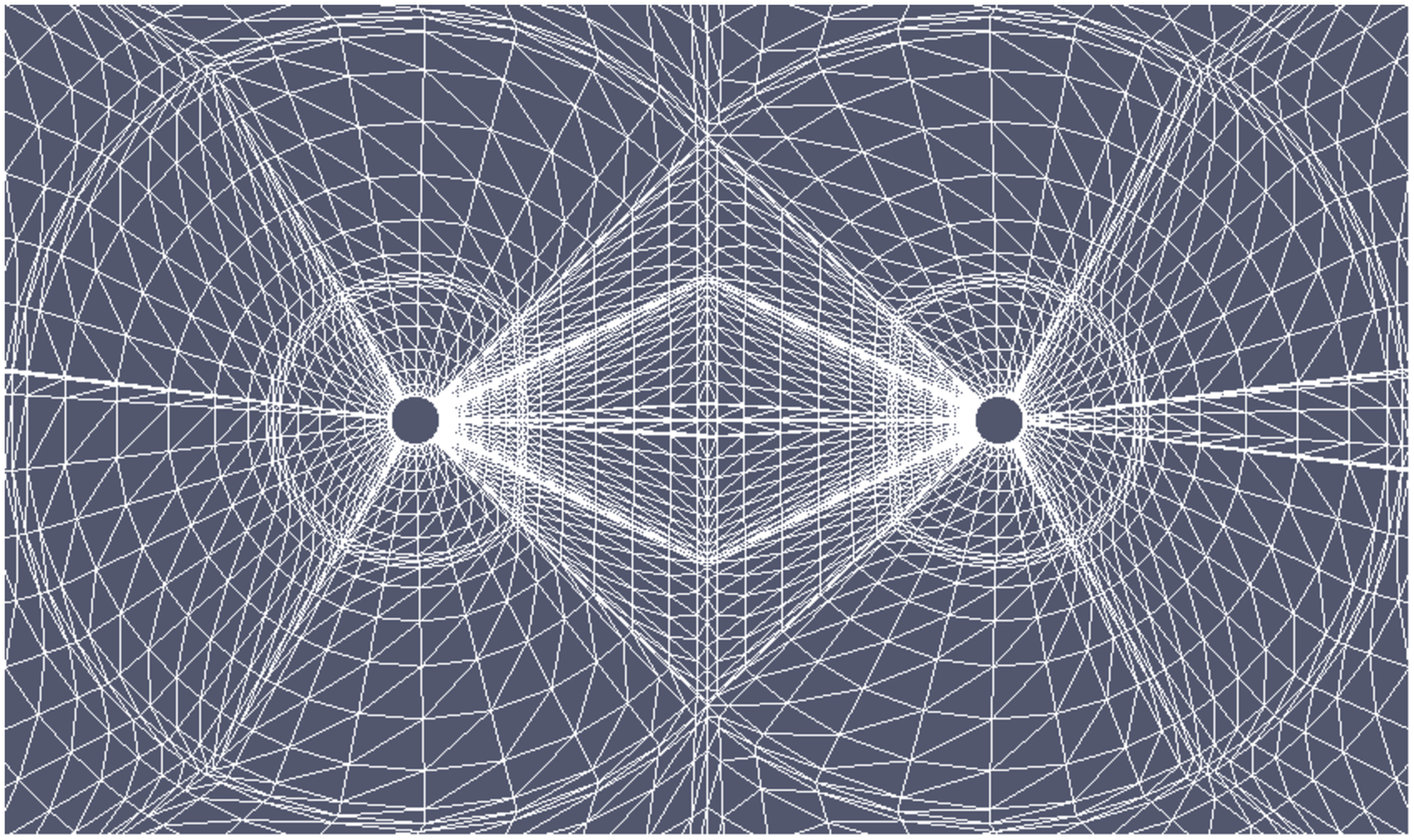}
  \end{center}
  \begin{center}
      \includegraphics[height=4cm]{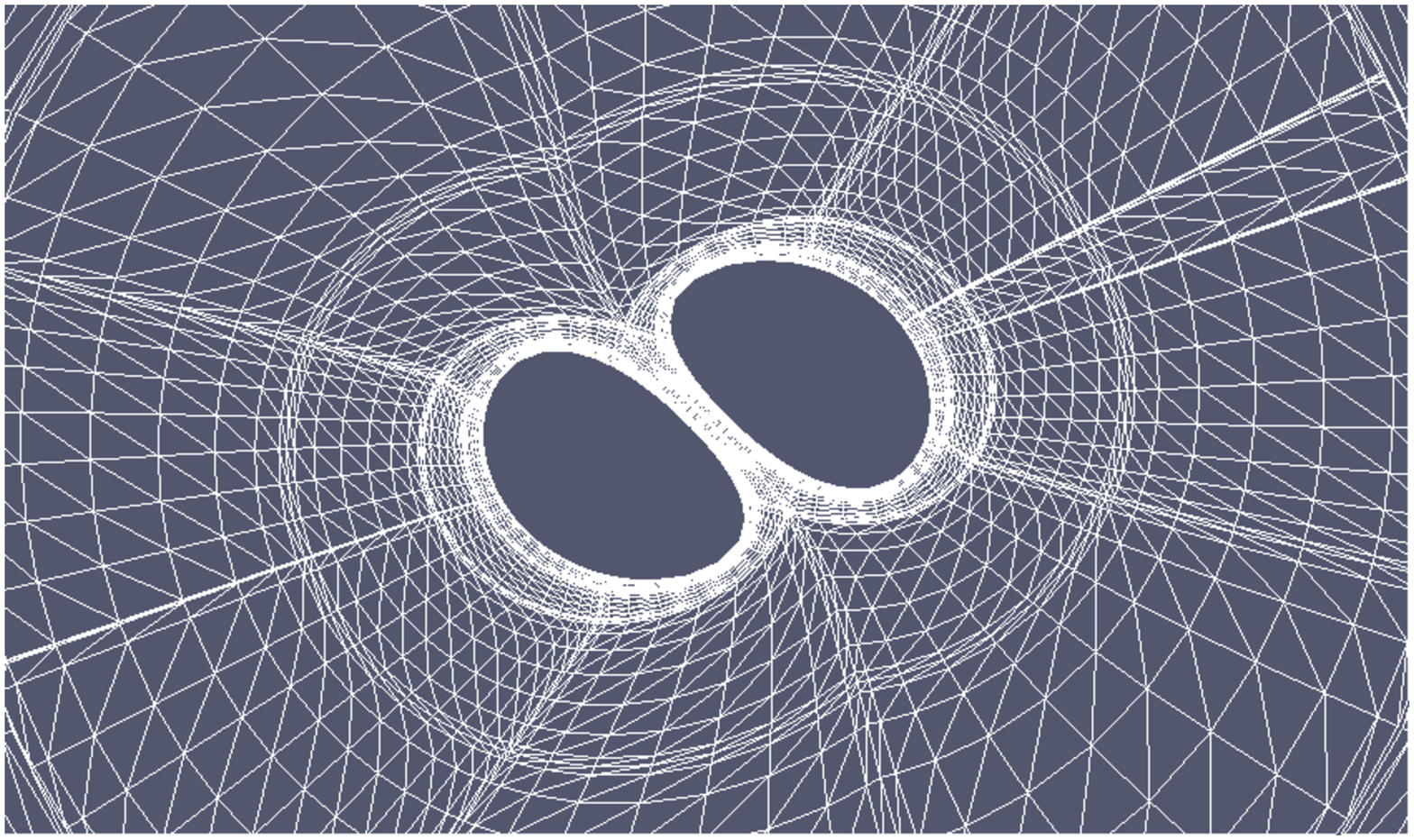}
  \end{center}
  \caption{a) Computational domain in grid coordinates; the black hole centers are at rest and the excision boundaries are spherical. 
b) Same domain in inertial coordinates near merger; the excision boundaries move and distort to track the apparent horizons.}
  \label{fig:domain}
\end{figure}

Control systems are used in SpEC for several purposes.  The most
important is their role in handling moving, excised black holes in a
spectral evolution method.  We use a dual-frame method~\cite{Scheel2006}
in which the grid is fixed in some coordinates $(t,x^i)$ but the components
of dynamical fields are expressed in a different coordinate system 
$(\inertial{t},\inertial{x}^i)$.  We call $(t,x^i)$ the \emph{grid coordinates}
and $(\inertial{t},\inertial{x}^i)$ the \emph{inertial coordinates}.
Figure~\ref{fig:domain} shows an example domain decomposition in both
coordinate systems.  
The two coordinate systems are connected
by a map $\cal M$ that depends on time-dependent parameters $\lambda(t)$.
The
excision boundaries are exactly spherical and stationary in grid coordinates.
In inertial coordinates, the apparent horizons move and distort as determined
by the solution of Einstein's equations supplemented by our gauge conditions.
The parameters $\lambda(t)$ need to be controlled so that the excision
boundaries in the inertial frame follow the motion and shapes of the
apparent horizons.  We use control systems to accomplish this.

The particular maps that we use will be discussed in 
Sec.~\ref{sec:control-systems-maps}.  In this section we will describe how
we construct the control system for a general parameter $\lambda(t)$, including
how we define the relationship between the control error $Q(t)$ and the 
control signal $U(t)$, how we smooth out noise in the control system, and 
how we dynamically adjust the feedback parameters and timescales.

\subsection{Definition of control errors and control signals}
\label{sec:ParameterPrediction}

We represent a general time-dependent map parameter $\lambda(t)$ as
a polynomial in time with a piecewise constant $N$th derivative:
\begin{equation}
\lambda(t) = 
\sum_{n=0}^{N} \frac{1}{n!}(t-t_i)^n\lambda^n_i, \qquad{\rm for}\quad
t_i \le t < t_{i+1},
\label{eq:GenericMapParam}
\end{equation}
where for each time interval $t_i \le t < t_{i+1}$ the
quantities $\lambda^n_i$ are constants.

At the beginning of
each new time interval $t_i$, we set the constants $\lambda^n_i$ in
 Eq.~\eqref{eq:GenericMapParam} as follows.  First for $n=N$ we set
$\lambda^N_i=U(t_i)$, where $U(t)$ is the control signal defined in
detail below. For $n<N$ we set
$\lambda^n_i=\left. d^n\lambda/dt^n\right|_{t=t_i}$, where the
derivative is evaluated at the end of the previous time interval.  In
this way all the derivatives of $\lambda(t)$ except the $N$th
derivative are continuous across intervals.  The goal will be to
compute the control signal $U(t)$ so as to drive the map parameter
$\lambda(t)$ to some desired behavior. Before we describe how to
compute the control signal $U(t)$, we first discuss the control error
$Q$, which will be used in the computation of $U(t)$.

To appropriately define the control error $Q$, one must answer the
question of how a small change in the map parameter corresponds to a
change in the observed variables.  If the control error is defined to
be too large, then the controller will consistently overshoot its
target, potentially leading to unstable behavior; conversely, if the
control error is defined to be too small, then the controller may
never be able to reach its target value.

If there exists a target value of $\lambda(t)$, call it
$\lambda\sub{target}$, that does not depend on the map but may
depend on other observable quantities in the system 
(call them $A$, $B$,\ldots), then we
define
\begin{equation}
Q = \lambda\sub{target}(A,B,\ldots) - \lambda.
\label{eq:Qgeneric}
\end{equation}
The goal is to drive $Q$ to zero and thereby drive $\lambda$ to
$\lambda\sub{target}$.  

If instead, as often happens for nonlinear systems, the target value
of $\lambda$ depends on $\lambda$ itself, even indirectly, then we 
define $Q$ differently using a generalization of
Eq.~\eqref{eq:Qgeneric}: we require that $\lambda$ attains its desired
value when $Q\to 0$, and we require that
\begin{equation}
\frac{\partial Q}{\partial \lambda} = -1 + {\cal O}(Q).
\label{eq:Qgenericnonlinear}
\end{equation}
The primary motivation for this condition is its anticipated use in 
relating time derivatives of $Q$ to those of $\lambda$ 
(see Eq.~\eqref{eq:dtqrelation}, below).
Note that in either Eqs.~\eqref{eq:Qgeneric} or~\eqref{eq:Qgenericnonlinear},
$Q$ could in principle be multiplied
by an arbitrary factor; however, if this were done, that factor would
need to be taken into account in the computation of the control
signal $U(t)$ below. Without loss of generality we assume no
additional scaling.  

In the case of several map parameters $\lambda_a(t)$ with
corresponding $Q_a$, where $a$ is an index that labels the map
parameters, the desired value of some $\lambda_a$ may depend on a
different map parameter $\lambda_b$.  In this case, we generalize
Eq.~\eqref{eq:Qgenericnonlinear} and require that each $Q_a$
satisfy
\begin{equation}
\frac{\partial{Q_a}}{\partial{\lambda_b}} = - \delta_{ab} + {\cal O}(Q),
\label{eq:Qdecoupled}
\end{equation}
where $\delta_{ab}$ is a Kronecker delta.  This criterion ensures that
we can treat each $\lambda_a$ independently when all control
errors are small. A way
of understanding Eq.~\eqref{eq:Qdecoupled} is to consider a set of
$Q'_a$ that are obtained via Eq.~\eqref{eq:Qgenericnonlinear} without
regard to coupling between different $\lambda_a$.  Then a set of $Q_a$
satisfying Eq.~\eqref{eq:Qdecoupled} can be obtained by diagonalizing
the matrix $\partial{Q'_a}/\partial{\lambda_b}$.  In the remainder of
this section we assume that if there are multiple map parameters
$\lambda_a$, the corresponding control errors $Q_a$ satisfy
Eq.~\eqref{eq:Qdecoupled}. We therefore drop the $a$ indices and
write equations for $U(t)$ in terms of a single $Q(t)$ satisfying
Eq.~\eqref{eq:Qgenericnonlinear} that represents the control error for
a single $\lambda(t)$.

The control error $Q(t)$ is a function of several observables. In the
case of the mapping functions that are designed to move and distort
the excision boundaries to follow the motion and shapes of the
apparent horizons, $Q(t)$ is some function of the current position or
shape of one or more apparent horizons.  The precise definition of
$Q(t)$ is different for each map parameter, and depends on the details
of how each map parameter couples to the observables.  We will discuss
the control error $Q$ for each of the map parameters in 
Sec.~\ref{sec:control-systems-maps}.
But it is not necessary to know the exact form of the control error 
in order to compute the control signal $U(t)$; it suffices to know only that
$U(t_i)$ is equal to $\lambda^N_i$ in Eq.~\eqref{eq:GenericMapParam},
and that the control error $Q$ obeys Eq.~(\ref{eq:Qgenericnonlinear}).

We now turn to the computation of the control signal $U(t)$.  
There is some flexibility in the control law determining $U(t)$,
so long as key feedback mechanisms are in place
(as shown in Sec.~\ref{sec:ControlTheory}).
In SpEC, we use either a standard PID controller,
\begin{equation}
\label{eq:pid} 
U(t) = a_0\int Q(t)\,dt + a_1\,Q(t) + a_2\ddt{Q},
\end{equation}
or a special PD (proportional-derivative) controller,
\begin{equation}
\label{eq:pnd} 
U(t) = \sum_{k=0}^K a_k\dtn{Q}{k},
\end{equation}
where typically $K=2$.

We set the constants $a_k$ so that the system damps $Q$ to zero
on some timescale $\tau_d$ that we choose.  We assume that $\tau_d$ is longer
than the interval $t_{i+1}-t_{i}$ defined in Eq.~\eqref{eq:GenericMapParam},
so that we can approximate $Q(t)$ and $\lambda(t)$ as smooth functions
rather than as functions with piecewise constant $N$th derivatives.
Under this assumption, we write
\begin{equation}
  U(t) = d^N\lambda/d t^N.
\label{eq:UequalsNthDerivOfLambda1}
\end{equation}
We also write
\begin{align}
\label{eq:dtqrelation}
  dQ/dt &= (\partial Q/\partial \lambda) (d\lambda/dt), \nonumber \\
        &= -d\lambda/dt.
\end{align}
In the first line
we have neglected the time dependence of other parameters besides 
$\lambda$
that enter into $Q$ under the assumption that the control system timescale
is shorter than the timescales of the quantities that we want to control.
In the second line we have used Eq.~\eqref{eq:Qgenericnonlinear} and
we have assumed that $Q$ is small.  Similarly
we write
\begin{align}
  d^2Q/dt^2 &= \frac{\partial Q}{\partial \lambda}  \frac{d^2\lambda}{dt^2}
              +\frac{\partial^2 Q}{\partial \lambda^2}
               \left(\frac{d\lambda}{dt}\right)^2, 
               \nonumber \\
            &= \frac{\partial Q}{\partial \lambda}  \frac{d^2\lambda}{dt^2},
               \nonumber \\
            &= -\frac{d^2\lambda}{dt^2},
\end{align}
where in the second line we have retained only terms of linear
order in $dQ/dt$ (and therefore in $d\lambda/dt$).  For higher derivatives
we continue to retain only terms linear in $Q$ and its derivatives,
so from Eq.~(\ref{eq:UequalsNthDerivOfLambda1}) we obtain
\begin{align}
U(t) &=  d^N\lambda/d t^N,\nonumber \\
     &= -d^NQ/d t^N.
\label{eq:UequalsNthDerivOfLambda}
\end{align}

For the PID controller and $N=2$, combining Eqs.~(\ref{eq:pid}) 
and~(\ref{eq:UequalsNthDerivOfLambda}) yields
\begin{equation}
\label{eq:pidode} 
-\frac{d^2 Q}{d t^2} = a_0\int Q(t)\,dt + a_1\,Q(t) + a_2\ddt{Q}.
\end{equation}
If we
choose $a_0=1/\tau_d^3$, $a_1=3/\tau_d^2$, and $a_2=3/\tau_d$, then the solution
to Eq.~\eqref{eq:pidode} will be exponentially damped on the timescale $\tau_d$,
\begin{equation}
Q\propto e^{-t/\tau_d}.
\end{equation}
The same exponential damping holds for the PD controller, Eq.~\eqref{eq:pnd},
for appropriate choices of the parameters $a_k$.  For $K=2$ and $N=3$, 
the parameters $a_k$ are identical to those in the PID case above.

\subsection{Averaging out noise}
\label{sec:averaging-out-noise}
The PID controller, Eq.~(\ref{eq:pid}), is computed by measuring the
control error $Q$, its time integral, and its time derivative.
The PD controller,
Eq.~(\ref{eq:pnd}), may require multiple derivatives of $Q(t)$
depending on the order $K$.  Generally only $Q$, and not its
derivatives or integrals, is available in the code at any given time
step. The simplest way to compute the derivatives of $Q$ is by finite
differencing in time, and the simplest way to compute the integral
is by a numerical quadrature.  

We measure the control error $Q$ at time intervals of length $\tau_m$,
where we choose $\tau_m < t_{i+1}-t_i$.  The measuring
time interval $\tau_m$ is then used as the time step for finite
difference stencils and quadratures.

The measured $Q$ is typically a function of apparent horizon locations
or shapes, and this measured $Q$ includes noise caused by the finite
resolution of the evolution, and the finite residual and finite number
of iterations of the apparent horizon finder.  Taking a numerical derivative of
$Q$ amplifies the noise, and then this noise
is transferred to the control signal via Eqs.~(\ref{eq:pid}) or~(\ref{eq:pnd}),
and then to the map.  If the noise amplitude is too large, the control
system will become unstable.  The PID controller generally handles
noise better than the PD controller for two reasons.  First, each
successive numerical derivative amplifies noise even further, so using
only one derivative instead of two (or more) results in a more
accurate control signal.  Second, the inclusion of an integral term
acts to further smooth the control signal.

In some cases, however, the noise in $Q$ can be problematic even for the PID
controller.  In these cases we implement direct averaging of the
control error in one of two ways: 1) we perform a polynomial fit of
order $N$ to the previous $M$ measurements of the control error, where
$M > N$, or 2) we perform an exponentially-weighted average, with
timescale $\tau_{\rm avg}$, of all previous control error measurements and their
derivatives and integrals.
The latter is our preferred method,
which we describe in detail in \ref{sec:ExpAverager}.

\subsection{Dynamic timescale adjustment}
\label{sec:DynamicTimescaleAdjustment}

In the previous section we have identified four timescales relevant
for each control system.  The first is the damping timescale $\tau_d$;
this describes how quickly the control error $Q(t)$ falls to zero, and
therefore how quickly the map parameter $\lambda(t)$ approaches its
desired value.  The second timescale is the control interval $\Delta
t_i := t_{i+1} - t_i$, which represents how often the $N$th
derivative of the function $\lambda(t)$ in
Eq.~(\ref{eq:GenericMapParam}) is updated.  The third is the
measurement timescale $\tau_m$, which indicates how often the
control error $Q$ 
is measured.  The fourth is the averaging timescale $\tau_{\rm avg}$,
which is used to smooth the control error $Q$ (and its derivatives
and integrals) for use in computing the control signal $U(t)$.
These timescales are not all independent; for example we have assumed
$\Delta t_i<\tau_d$ in deriving Eq.~(\ref{eq:pidode}), and we have
assumed $\tau_m<\Delta t_i$ so that we can obtain smooth measurements
of the derivatives of $Q$.  

Because binary black hole evolutions are nonlinear dynamic systems,
we adjust the damping timescale, $\tau_d$, throughout the simulation.
We then set the three other timescales, $\Delta t_i$, $\tau_m$, and
$\tau_{\rm avg}$, based on the current value of the damping timescale
$\tau_d$, as we now describe.

The timescale $\tau_d$ should be shorter than the timescale on which
the physical system changes; otherwise, the control system cannot
adjust the map parameters quickly enough to respond to changes in the
system.  But if the timescale $\tau_d$ is too small, then the
measurement timescale $\tau_m$ on which we compute the apparent
horizon must also be small, meaning that frequent apparent horizon
computations are needed; this is undesirable because computing
apparent horizons is computationally expensive.  We would like to
adjust $\tau_d$ in an automatic way so that it is relatively large
during the binary black hole inspiral, decreases during the plunge and merger, and
increases again as the remnant black hole rings down.  In the
canonical language of control theory, we would like to implement
``gain scheduling''~\cite{Sontag1998},
tuning the behavior of the control system for
different operating regimes.

We do this as follows: For all map parameters (except the $\ell=0$
component of the horizon shape map, which is treated differently; see
Sec.~\ref{sec-AdaptiveSizeControl}), the damping timescale is a generic
function of $Q$ and $\dot{Q}$, i.e., the error in its associated map
parameter and its derivative.  Whenever we adjust the control
signal $U(t)$ at interval $t_i$, 
we also tune the timescale in the following way:
\begin{equation}
\tau_d^{i+1} = \beta \tau_d^i,
\end{equation}
where typically
\begin{equation}
\beta = \left\{
  \begin{array}{ll}
    0.99, & {\rm ~if~} \dot{Q}/Q > -1/2\tau_d {\rm ~and~} |Q| {\rm ~or~} |\dot{Q}\tau_d| > Q_t^{\rm Max} \\
    1.01, & {\rm ~if~} |Q| < Q_t^{\rm Min} {\rm ~and~} |\dot{Q}\tau_d| < Q_t^{\rm Min} \\
    1, & {\rm ~otherwise.}
  \end{array}\right.
\end{equation}
Here $Q_t^{\rm Min}$ and $Q_t^{\rm Max}$ are thresholds for the control
error $Q$.  The idea is to keep $|Q|<Q_t^{\rm Max}$ so that the
map parameters are close to their desired values, but to also keep
$|Q|>Q_t^{\rm Min}$ because an unnecessarily small $|Q|$ 
means unnecessarily small timescales and therefore a large
computational expense (because the apparent horizon must be found frequently).
For binary black hole simulations where the holes have masses $M_A$ and
$M_B$, we find that the following choices work well:
\begin{align}
Q_t^{\rm Max} &= \frac{2\xten{-3}}{M_A/M_B + M_B/M_A} \\
Q_t^{\rm Min} &= \frac{1}{4}Q_t^{\rm Max}.
\end{align}

Once we have adjusted the timescale $\tau_d$ for every control system,
we then use these timescales to choose the times $t_{i+1}$ in
Eq.~\eqref{eq:GenericMapParam} at which we update the polynomial coefficients
$\lambda_i^n$ in the map parameter expression $\lambda(t)$,
\begin{equation}
\Delta t_i := t_{i+1} - t_i = \alpha_d\min(\tau_d),
\end{equation}
where typically $\alpha_d=0.3$, and the minimum is taken over all
map parameters (except for the $\ell=0$ component of the horizon
shape map).  This ensures that the coefficients $\lambda_i^n$, are 
updated faster than the physical system is changing, and faster than
the control system is damping.  For $\alpha_d$ too large, we find that
the control system becomes unstable.

We also use the timescale $\tau_d$ to choose the interval $\tau_m$
at which we measure the control error. 
For many map parameters, the associated control
error is a function of apparent horizon quantities, 
which is why we desire $\tau_m$ to be as large as possible.  
But a certain number of measurements are needed for each 
$\Delta t_i$ so that the
control signals, defined in Eqs.~\eqref{eq:pid} and~\eqref{eq:pnd},
are sufficiently accurate and the control system is stable. We choose
\begin{equation}
\tau_m = \alpha_m \Delta t_i,
\end{equation}
where $\alpha_m$ is typically between $0.25$ and $0.3$. In other
words, we measure the control error three or four times before we
update the control signal.
This also ensures that the
averaging timescale is greater than the measurement
  timescale, as we typically
  choose $\tau_{\rm avg}\sim 0.25 \tau_{d}$.

\section{Control Systems for Maps}
\label{sec:control-systems-maps}
In a SpEC evolution, we transform the grid coordinates, $\grid{x}^i$,
into inertial coordinates, $\inertial{x}^i$, through a series of
elementary maps as
in~\cite{Scheel2006,Boyle2007,Scheel2009,Szilagyi:2009qz,Buchman:2012dw}.
Several maps have been added and many improvements have been made
since their original introduction. The full transformation is
$\inertial{x}^i = \mathcal{M} \grid{x}^i$, where
\begin{equation}
  \label{eq:MapSequence}
\begin{array}{ll}
\mathcal{M} =&
\map{Translation}
\circ\map{Rotation}
\circ\map{Scaling}  \\
& \circ\map{Skew}
\circ\map{CutX}
\circ\map{Shape}.
\end{array}
\end{equation}
Below we will describe each of these maps and how we measure the error
in their parameters.  Before we do this, however, we describe our
domain decomposition and how we measure apparent horizons, because information
from the grid and the horizons is used to determine the maps.

In the grid coordinates $\grid{x}^i$, the domain decomposition looks
like Fig.~\ref{fig:CutSphere}. There are two excision boundaries, $A$
and $B$, which are spheres in grid coordinates.  The grid-coordinate
centers of these excision boundaries we will call $C^i_H$, where $H$
is either $A$ or $B$.  The excision boundaries (and therefore $C^i_H$)
remain fixed throughout the evolution.  The purpose of many of the
maps is to move the {\em mapped} centers 
$\inertial{C}^i_H := \mathcal{M}(C^i_H)$ along with the
centers of the apparent horizons.

We measure the apparent horizons in an intermediate frame
$\distorted{x}^i$,
which we call the \emph{distorted frame}.
This frame is connected to the grid frame by the map
\begin{equation}
\map{Distortion} = \map{CutX}\circ\map{Shape}.
\label{eq:DistortionMapSymbol}
\end{equation}
We will discuss exactly what \map{Shape} and
\map{CutX} are below, but for now we need only demand
that \map{Distortion} has two properties:
The first is that it leaves the centers of the excision boundaries invariant,
i.e., 
\begin{equation}
\distorted{C}^i_H := \map{Distortion}(C^i_H) = C^i_H.
\label{eq:DistortionMapPreservesCenters}
\end{equation}
The second property is that at each excision boundary $H$, 
\map{Distortion} leaves angles invariant. That
is, if we define grid-frame polar coordinates
$(r_H,\theta_H,\phi_H)$ centered about excised region $H$
in the usual way,
\begin{align}
  x^0        &= C_H^0 + r_H \sin\theta_H \cos\phi_H, \label{eq:PolarCoords1} \\
  x^1        &= C_H^1 + r_H \sin\theta_H \sin\phi_H, \label{eq:PolarCoords2} \\
  x^2        &= C_H^2 + r_H \cos\theta_H,            \label{eq:PolarCoords3}
\end{align}
and similarly for polar coordinates
$(\distorted{r}_H,\distorted{\theta}_H,\distorted{\phi}_H)$ centered about
$\distorted{C}^i_H$ in the $\distorted{x}^i$ frame,
then the second property means that
\begin{equation}
\begin{array}{rl}
\distorted{\theta}_H &=\theta_H,\\
\distorted{\phi}_H   &=\phi_H,
\end{array}\label{eq:DistortionMapPreservesAngles}
\end{equation} 
on the excision boundary. Note that the index $i$ on spatial
coordinates ranges over $(0,1,2)$ in this paper.

We find the apparent horizons in the frame $\distorted{x}^i$ 
by a fast flow method~\cite{Gundlach1998}.  Each horizon is
represented as a smooth surface in this frame, and is parameterized
in terms of polar coordinates centered around $\distorted{C}^i_H$:
the radius of the horizon at each $(\distorted{\theta}_H,\distorted{\phi}_H)$
is given by
\begin{equation}
\distorted{r}^{\rm AH}_H(\distorted{\theta}_H,\distorted{\phi}_H) = \sum_{\ell m} 
\distorted{S}^H_{\ell m} Y_{\ell m}(\distorted{\theta}_H, \distorted{\phi}_H),
\label{eq:StrahlkorperAH}
\end{equation}
where again the index $H$ labels which excision boundary is enclosed by
the apparent horizon.  Here 
$Y_{\ell m}(\distorted{\theta}_H,\distorted{\phi}_H)$ 
are spherical harmonics and $\distorted{S}^H_{\ell m}$ are expansion
coefficients that describe the shape of the apparent horizon.

We search for apparent horizons in the distorted frame 
$\distorted{x}^i$ rather than in the grid frame $x^i$ or the
inertial frame $\inertial{x}^i$ because this simplifies the
formulae for the control systems. In particular, this choice decouples
the control errors of the shape map $Q_{\ell m}$
(defined in Sec.~\ref{sec:shape-control}, Eq.~(\ref{eq:Qlm})) 
for different values of $(\ell,m)$, and
it decouples the errors $Q_{\ell m}$ from the control errors of 
the maps connecting the distorted and inertial frames.

For each surface $H$, we
define the center of the apparent horizon
\begin{equation}
  \label{eq:CenterAhIntegral}
  \distorted{\xi}^i_H = 
  \frac{\int_H \distorted{x}^i 
    (\distorted{r}_H^{\rm AH})^2 d\distorted\Omega}
       {\int_H (\distorted{r}_H^{\rm AH})^2 d\distorted\Omega},
\end{equation}
where the integrals are over the surface.
In the code, it suffices to 
use the following approximation of Eq.~(\ref{eq:CenterAhIntegral}),
which becomes
exact as the surface becomes spherical:
\begin{align}
  \distorted{\xi}^0_H &= C^0_H-\sqrt{3/2\pi}
  \,\Re(\distorted{S}^H_{11}),\label{eq:CenterAhx} \\
  \distorted{\xi}^1_H &= C^1_H+\sqrt{3/2\pi}
  \,\Im(\distorted{S}^H_{11}),\label{eq:CenterAhy} \\
  \distorted{\xi}^2_H &= C^2_H+\sqrt{3/4\pi}
  \,\distorted{S}^H_{10}.\label{eq:CenterAhz}
\end{align}
Here we have used the property $\distorted{C}^i_H = C^i_H$, 
Eq.~\eqref{eq:DistortionMapPreservesCenters}.
We distinguish the center $C^i_H$ of the excision boundary
from the center $\distorted{\xi}^i_H$ of the corresponding apparent horizon.
The former is fixed in time in the grid frame, but the latter will change as
the metric quantities evolve.  The purpose of several of the maps 
(namely \map{Scaling}, \map{Rotation},
and \map{Translation} described below)
is to ensure that $\distorted{\xi}^i_H-C^i_H$ is driven toward
zero; i.e., to ensure 
that the centers of the excision boundaries track the centers
of the apparent horizons.

We now describe each of the maps comprising $\mathcal{M}$
and how we measure the error in their parameters.
The order in which we discuss the maps is not the same as the
order in which the maps are composed so that we can discuss
the simplest maps first.
To produce less cluttered equations in the following
descriptions of the maps, we omit accents on variables that represent specific
frames (i.e. we just write $x$ instead of $\distorted{x}$
or $\inertial{x}$) whenever the input or output frame of the map
is unambiguous or explicitly stated.

\subsection{Scaling}
\label{sec:Scaling}
The scaling map, \map{Scaling}, causes the grid to
shrink or expand (and the
excision boundaries to move respectively closer together 
or farther apart) in the inertial frame,
thereby allowing the grid to follow the two black holes as their 
separation changes. 

This map transforms the radial coordinate
with respect to the origin (i.e., the center of the outermost sphere in
Fig.~\ref{fig:CutSphere}), such that the region near the black holes
is scaled uniformly by a factor $a$ and the outer boundary is scaled 
by a factor $b$,
\begin{equation}
  R \mapsto aR + (b-a) R^3/R^2_{\rm OB}.
\label{eq:ExpansionMap}
\end{equation}
Here $R$ is the radial coordinate, $R_{\rm OB}$ is the radius
at the outer boundary, $a$ is a parameter that will be determined
by a control system, and $b$ is another parameter that will be
determined empirically.

In~\cite{Scheel2006}, 
the scaling map was simply $x^i \mapsto a(t) x^i$, which is
recovered by Eq.~\eqref{eq:ExpansionMap} if $b=a$.  In that
case, as the black holes inspiral together and $a$ decreases, the outer
boundary of the grid decreases as well.  For long evolutions,
the outer boundary
decreases so much that we can no longer extract gravitational radiation far 
from the hole.  The addition of $b$ to the map alleviates this difficulty,
allowing the motion of the outer boundary to be decoupled from the motion 
of the holes.  We choose $b$ by an explicit functional form 
\begin{equation}
\label{eq:ScalingMapb}
  b(t) = 1-10^{-6} t^3/(2500+t^2),
\end{equation}
which is designed to keep the outer boundary from shrinking
rapidly, but to allow the boundary to move inward at a small speed,
so that zero-speed modes are advected off the grid (and thus need no
boundary condition imposed on them). 
We choose a cubic function of
time in Eq.~\eqref{eq:ScalingMapb} because we have 
found that at least the first two time derivatives of $b(t)$ 
must vanish initially, or else a significant ingoing pulse of constraint 
violations is produced at the outer boundary.

We control the scale factor $a$ of this map using the error function
\begin{equation}
Q_a = a(dx^0 - 1),
\label{eq:Qa}
\end{equation}
where
\begin{equation}
dx^i = \frac{\distorted{\xi}^i_A-\distorted{\xi}^i_B}{C^0_A - C^0_B}.
\label{eq:dx}
\end{equation}
We assume that the separation
vector
$C^i_A-C^i_B$ in the grid frame is parallel to the
$x$-axis.
  The idea is that the distorted-frame separation of the
horizon
centers
along the $x$-axis
is
driven to be the same as the separation of
the excision boundary centers.

To show that the control system for $a$ obeys Eq.~(\ref{eq:Qgenericnonlinear}),
we consider the change in $Q_a$ under variations of $a$ with
all other maps held fixed, and with the inertial-coordinate
centers $\inertial{\xi}^i_H$ of the horizons held 
fixed.\footnote{
We consider inertial-frame quantities to be 
fundamental
and determined by the solution of Einstein's equations plus gauge
conditions, and therefore independent (modulo numerical truncation
error) of the numerical grid and of the maps 
that we use to construct, move, and distort that grid.
}
This means that
the distorted-frame centers of
the horizons $\distorted{\xi}^i_H$ appearing in Eq.~(\ref{eq:dx}) change
with variations of $a$.  The second term in Eq.~(\ref{eq:ExpansionMap})
is small when evaluated near the horizon where $(R/R_{\rm OB})^2 \ll 1$, so we
can write the action of the scaling map as
\begin{equation}
  \label{eq:ScalingMapActionOnCenters}
  \inertial{\xi}^i_H = a\distorted{\xi}^i_H,
\end{equation}
and therefore under variations $\delta a$,
\begin{equation}
  \label{eq:ScalingMapActionOnCentersVar}
  0=\delta \inertial{\xi}^i_H = 
  \distorted{\xi}^i_H \,\delta a + a\,\delta\distorted{\xi}^i_H.
\end{equation}
Taking variations of Eq.~(\ref{eq:Qa}), 
using Eq.~(\ref{eq:ScalingMapActionOnCentersVar}) 
to substitute for $\delta\distorted{\xi}^i_H$, and noting that the
excision-boundary centers $C^i_H$ are constants and do not vary 
with $a$, we obtain
\begin{equation}
  \label{eq:QaVariation}
  \delta Q_a = -\delta a,
\end{equation}
so that Eq.~(\ref{eq:Qgenericnonlinear}) is satisfied.

  To verify that the more general decoupling equation,
  Eq.~(\ref{eq:Qdecoupled}), is satisfied for the scaling map, we must
  show, in addition to Eq.~(\ref{eq:QaVariation}), that the quantity
  $dx^0$ defined in Eq.~(\ref{eq:dx}) is invariant under the action of
  the other maps, at least in the limit that all the control errors
  $Q$ are small.  We consider each map in turn.
  The translation map moves both apparent horizons
  together, so it leaves $dx^0$ invariant.  Changes in the rotation
  map parameters will change $dx^0$ only by an amount proportional to the
  control errors of the rotation map (see
  Eqs.~(\ref{eq:RotationVariationPhi})
  and~(\ref{eq:RotationVariationTheta}) below).  The skew map (below) leaves
  the centers of the excision boundaries invariant. Because the intent
  of the rotation, translation, and scaling maps is to drive the
  centers of the apparent horizons toward the centers of the excision
  boundaries, this means that the skew map changes $dx^0$ by an amount 
  proportional to the control errors of the rotation, translation, and scaling
  maps.  Finally, the shape and CutX maps connect the distorted
  and grid frames, so they cannot affect $dx^0$.

\subsection{Rotation}
The rotation map, \map{Rotation}, is a
rigid 3D rotation about the origin that
tracks the orbital phase and precession of the system,
\begin{equation}
\label{eq:RotationMap}
x^i \mapsto \mathcal{R}^i_j x^j, \;\;\; \mbox{where} \;\;
\mathcal{R} = \left(
\begin{array}{ccc}
\cos\vartheta\cos\varphi & -\sin\vartheta & \cos\vartheta\sin\varphi \\
\sin\vartheta\cos\varphi & \cos\vartheta  & \sin\vartheta\sin\varphi \\
-\sin\varphi & 0 & \cos\varphi
\end{array}\right).
\end{equation}
The pitch and yaw map parameters $(\varphi, \vartheta)$ are controlled
so as to align the line segment connecting the apparent horizon centers
with the distorted-frame $x$-axis.
Note that the map parameters $(\varphi, \vartheta)$
are functions of time, and are not to be confused with the polar coordinates 
$(\phi_H,\theta_H)$ centered about each excision boundary.
Then the control error is given by
\begin{align}
\label{eq:Qphi}
Q_{\varphi}   &= -\frac{\distorted{\xi}^2_A-\distorted{\xi}^2_B}{\distorted{\xi}^0_A-\distorted{\xi}^0_B} \\
\label{eq:Qtheta}
Q_{\vartheta} &=  \frac{\distorted{\xi}^1_A-\distorted{\xi}^1_B}{(\distorted{\xi}^0_A-\distorted{\xi}^0_B) \cos\varphi}.
\end{align}
In the case of extreme precession where $\varphi\rightarrow\pi/2$, these equations
are insufficient because $Q_{\vartheta}$ diverges. Our solution is to use
quaternions, which avoid this singularity (for a complete discussion,
see~\cite{Ossokine:2013zga}).

The control errors $Q_{\vartheta}$ and $Q_{\varphi}$ obey
  Eq.~(\ref{eq:Qdecoupled}).  To show this, consider variations of
  $\vartheta$ and $\varphi$ with other maps held fixed, and with the
  inertial-coordinate centers $\inertial{\xi}^i_H$ of the horizons
  held fixed. The rotation map, Eq.~(\ref{eq:RotationMap}), implies that
  under these variations,
\begin{equation}
  \delta \inertial{\xi}^i_H = 0 = {\cal R}^i_j \delta   \distorted{\xi}^j_H 
                                + (\delta {\cal R})^i_j \distorted{\xi}^j_H.
\end{equation}
Multiplying this equation by ${\cal R}^{-1}$ we obtain
\begin{equation}
  \delta \distorted{\xi}^i_H = -({\cal R}^{-1} \delta {\cal R})^i_j 
  \distorted{\xi}^j_H,
\end{equation}
which yields
\begin{align}
  \label{eq:RotationVariationPhi}
  \frac{\partial \distorted{\xi}^i_H}{\partial \varphi} &= 
   \mathcal{A}^i_j\distorted{\xi}^j_H, \;\;\; \mbox{where} \;\;
  \mathcal{A} := -\mathcal{R}^{-1}\frac{\partial\mathcal{R}}{\partial\varphi} =
  \left(\begin{array}{rrr}
    0&0&-1\\
    0&0&0\\
    ~1&~0&0
  \end{array}\right) \\
  \label{eq:RotationVariationTheta}
  \frac{\partial \distorted{\xi}^i_H}{\partial \vartheta} &= 
    \mathcal{B}^i_j\distorted{\xi}^j_H, \;\;\; \mbox{where} \;\;
  \mathcal{B} := -\mathcal{R}^{-1}\frac{\partial\mathcal{R}}{\partial\vartheta}=
  \left(\begin{array}{ccc}
    0&\cos\varphi&0\\
    -\cos\varphi &0&-\sin\varphi\\
    0&\sin\varphi &0
  \end{array}\right).
\end{align}
We can now verify Eq.~(\ref{eq:Qdecoupled}) for the special case where
the indices $a$ and $b$ in Eq.~(\ref{eq:Qdecoupled}) are either
$\vartheta$ and $\varphi$.  This result is obtained in a straightforward
way by differentiating Eqs.~(\ref{eq:Qphi}) 
or~(\ref{eq:Qtheta}) with respect to $\vartheta$ or $\varphi$, 
and substituting Eqs.~(\ref{eq:RotationVariationPhi}) 
or~(\ref{eq:RotationVariationTheta}).

In addition, the control errors $Q_{\varphi}$ and $Q_\vartheta$
are independent (to leading order in the control errors) 
of changes in the parameters of the other maps that
make up $\cal M$:  Variations of Eqs.~(\ref{eq:Qphi})
and~(\ref{eq:Qtheta}) with respect to the scaling map parameter $a$
are zero, because both the numerator and denominator of $Q_{\varphi}$
and $Q_\vartheta$ scale in the same way with $a$.  Similarly,
$Q_\varphi$ and $Q_\vartheta$ are independent of the translation map,
since both apparent horizons are translated by the same amount.  The
skew map can change $Q_{\varphi}$ and $Q_\vartheta$, but only by an
amount proportional to control errors, because the skew map leaves
$C^i_H$ invariant and other maps ensure that $\distorted{\xi}^i_H$ are
close (within a control error) to $C^i_H$. The shape and CutX maps
cannot affect $Q_{\varphi}$ and $Q_\vartheta$ because those maps
connect the grid and distorted frames (and therefore they cannot
  change the distorted-frame horizon centers $\distorted{\xi}^i_H$).

\subsection{Translation}
\label{sec:translation}
The translation map, \map{Translation}, transforms the Cartesian
coordinates, $x^i$, according to
\begin{equation}
  x^i \mapsto x^i + f(R) T^i,
\label{eq:TranslationMap}
\end{equation}
where $f(R)$ is a Gaussian centered on the origin with a width set such 
that $f(R)$ falls off to machine precision at the outer
boundary radius,
and $T^i(t)$ are translation parameters that are 
adjusted by a control system.

The translation map moves the grid to account for any drift of the
``center of mass'' of the system (as computed assuming point masses at
the apparent horizon centers) in the inertial frame.
This drift can be caused by momentum
exchange between the black holes and the surrounding gravitational
field~\cite{Chen2009,Lovelace:2009}, by anisotropic radiation of
linear momentum to infinity, or by linear momentum in the initial data.
This is the third map (the other two
are rotation and scaling) that drives the centers of the apparent horizons 
toward the centers of the excision boundaries. The apparent horizon
centers $\distorted{\xi}^i_A$ and $\distorted{\xi}^i_B$ represent
six degrees of freedom:
one is fixed by the scaling map, two by rotation, and three
by translation.

The control errors will be more complicated than for the other control systems
because translation and rotation do not commute.  
We define the control errors $Q_T^i$ for each of the translation directions
$i=0,1,2$ as
\begin{equation}
  \label{eq:QTranslation0}
  Q_T^i = a {\cal R}^i_j \left[\distorted{\xi}^j_B + {\cal P}^j_k
                         (\distorted{\xi}^k_A-\distorted{\xi}^k_B)\right],
\end{equation}
where ${\cal R}$ is the rotation matrix in Eq.~(\ref{eq:RotationMap}), 
and ${\cal P}$ is some matrix yet to be determined, which may depend on
the rotation parameters $(\varphi,\vartheta)$ and on the constants $C^i_H$, 
but which may not depend on the translation parameters $T^i$.
This control error must have the property that 
$Q_T^i=0$ when $\distorted{\xi}^i_H=C^i_H$, so our first
restriction on ${\cal P}$ is that it satisfies
\begin{equation}
  \label{eq:TranslationQZeroRequirement}
  0 = C^i_B + {\cal P}^i_j(C^j_A-C^j_B).
\end{equation}

To check Eq.~(\ref{eq:Qdecoupled}), we note that near the black holes
we can neglect the last term in Eq.~(\ref{eq:ExpansionMap}) and we
can use $f(R)\sim 1$ in Eq.~(\ref{eq:TranslationMap}), so that the apparent
horizon centers in the inertial and distorted frames are related by
\begin{equation}
  \label{eq:KinematicalInverseMapNearAH}
  a\mathcal{R}^i_j\distorted{\xi}^j_H = \inertial{\xi}^i_H -  T^i.
\end{equation}
Inserting Eq.~(\ref{eq:KinematicalInverseMapNearAH}) 
into Eq.~(\ref{eq:QTranslation0}), we obtain
\begin{equation}
  \label{eq:QTranslation1}
  Q_T^i = \inertial{\xi}^i_B - T^i - ({\cal R}{\cal P}{\cal R}^{-1})^i_j
  (\inertial{\xi}^j_A-\inertial{\xi}^j_B).
\end{equation}
The second term in Eq.~(\ref{eq:QTranslation1}) is the only term that depends
on the translation parameter $T^i$, so we have
\begin{equation}
  \frac{\partial Q_T^i}{\partial T^j} = -\delta_{ij}
\end{equation}
When varying map parameters, the inertial-frame horizon centers remain fixed,
so the only other term in Eq.~(\ref{eq:QTranslation1}) that depends on
map parameters is the last term, which depends on $(\vartheta,\varphi)$ 
because of the rotation matrices and because of the $(\vartheta,\varphi)$
dependence in ${\cal P}$.
We can therefore write the variation of $Q_T^i$ with respect 
to $(\vartheta,\varphi)$ as
\begin{align}
\frac{\partial Q_T^i}{\partial W} 
&= -\frac{\partial}{\partial W}({\cal R}{\cal P}{\cal R}^{-1})^i_j
(\inertial{\xi}^j_A-\inertial{\xi}^j_B),\\
\label{eq:dQTranslationdW}
&=  -\frac{\partial}{\partial W}({\cal R}{\cal P}{\cal R}^{-1})^i_j a {\cal R}^j_k
 (\distorted{\xi}^k_A-\distorted{\xi}^k_B),
\end{align}
where $W$ stands for either $\vartheta$ or $\varphi$, and where
in the last line we have used Eq.~(\ref{eq:KinematicalInverseMapNearAH}).
To obey Eq.~(\ref{eq:Qdecoupled}), $\partial Q_T^i/\partial W$ must either be
zero, or on the order of a control error $Q$.
For $i=1,2$, the quantity $(\distorted{\xi}^i_A-\distorted{\xi}^i_B)$ 
is proportional to the control error of the rotation map, Eqs.~(\ref{eq:Qphi})
and~(\ref{eq:Qtheta}), so these terms can be neglected
in Eq.~(\ref{eq:dQTranslationdW}). However, 
for $i=0$, $(\distorted{\xi}^i_A-\distorted{\xi}^i_B)$ is not proportional 
to a control error; instead 
$(\distorted{\xi}^0_A-\distorted{\xi}^0_B)$ is driven to a constant finite
value of $C^0_A-C^0_B$ by the scaling control system.  Therefore, in order
to satisfy Eq.~(\ref{eq:Qdecoupled}), we require
\begin{equation}
\label{eq:TranslationDecouplingRequirement}
  -\frac{\partial}{\partial W}({\cal R}{\cal P}{\cal R}^{-1}){\cal R} \left(
  \begin{array}{c}
    1\\
    0\\
    0
   \end{array}
  \right) = 0.
\end{equation}
We find that we can satisfy 
both Eqs.~(\ref{eq:TranslationQZeroRequirement}) 
and~(\ref{eq:TranslationDecouplingRequirement}) by choosing
\begin{equation}
  {\cal P} = 
  \frac{1}{C^0_B-C^0_A}
  \left(
  \begin{array}{ccc}
     C^0_B & -C^1_B & -C^2_B\\
     C^1_B & C^0_B + C^2_B \tan\varphi & 0           \\
     C^2_B & -C^1_B \tan\varphi & C^0_B 
  \end{array}
  \right).  
\end{equation}
Here we have again assumed that the separation between the centers of
the excision boundaries is parallel to the $x$-axis,
i.e., $C^1_A=C^1_B$ and $C^2_A=C^2_B$.

\subsection{Skew}
\label{sec:skew}
\begin{figure}
\centerline{\includegraphics[scale=0.37]{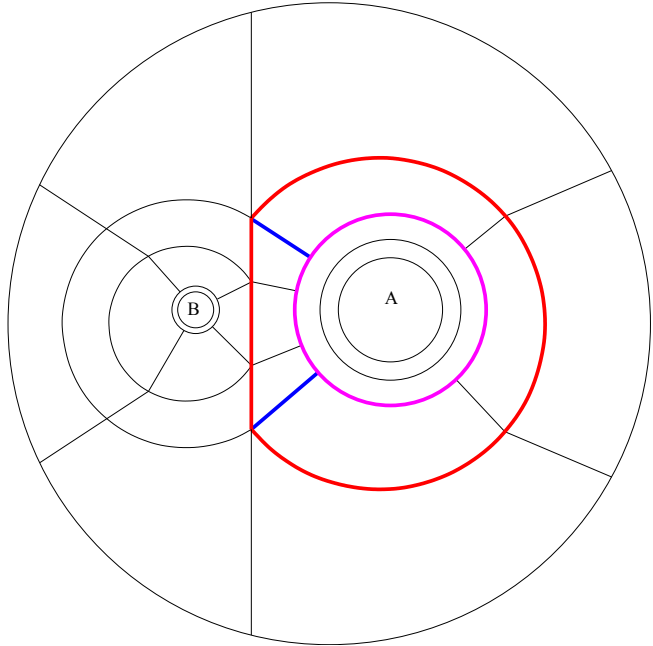}}
\caption{\label{fig:CutSphere} Two-dimensional projection of the
  domain decomposition near the two
  black holes A and B.  Shown are boundaries between subdomains.  Each subdomain
  takes the shape of a spherical shell, a distorted cylindrical shell, or a
  distorted cylinder.  Additional spherical-shell subdomains (not shown) 
  surround the outer boundary of the figure and extend to large radius.
  This domain decomposition is explained in detail in 
  the Appendix of~\cite{Buchman:2012dw}.
  The red, magenta, and blue surfaces are those for
  which the function $f_A(r_A,\theta_A,\phi_A)$ from 
  Eq.~\eqref{eq:GridToShape} has a discontinuous gradient,
  as described in Sec.~\ref{sec:shape-control}.}
\end{figure}

In Fig.~\ref{fig:CutSphere}, a prominent feature of the domain
decomposition is a plane (a vertical line in the two-dimensional
figure) that is perpendicular to the grid-frame $x$-axis and lies
between the two excision boundaries $A$ and $B$.  
We call this plane the ``cutting plane''. 

\begin{figure}
  \begin{minipage}{.5\textwidth}
  \begin{flushright}
      \includegraphics[width=4cm]{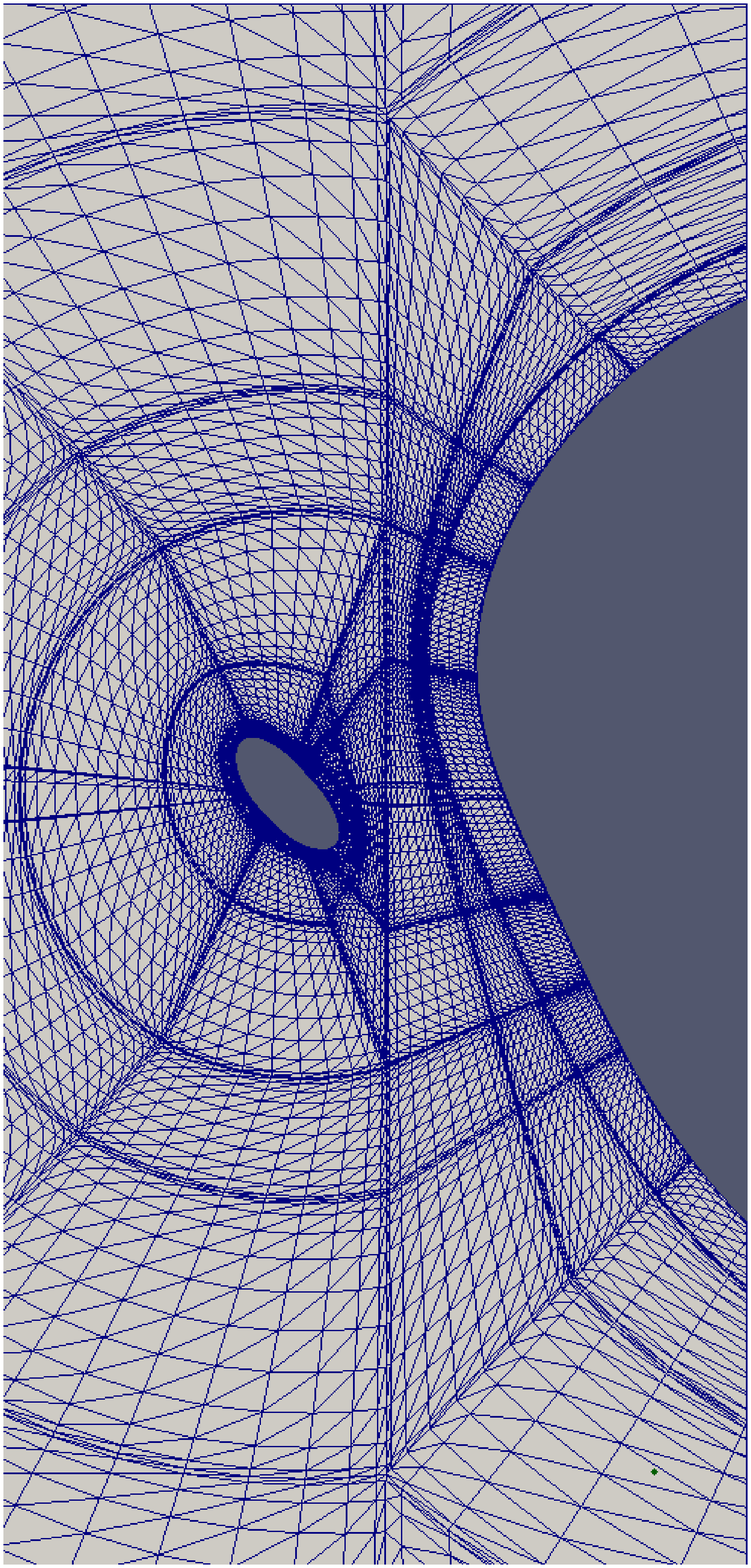}
      \,
  \end{flushright}
  \end{minipage}
  \begin{minipage}{.5\textwidth}
  \begin{flushleft}
      \,
      \includegraphics[width=4cm]{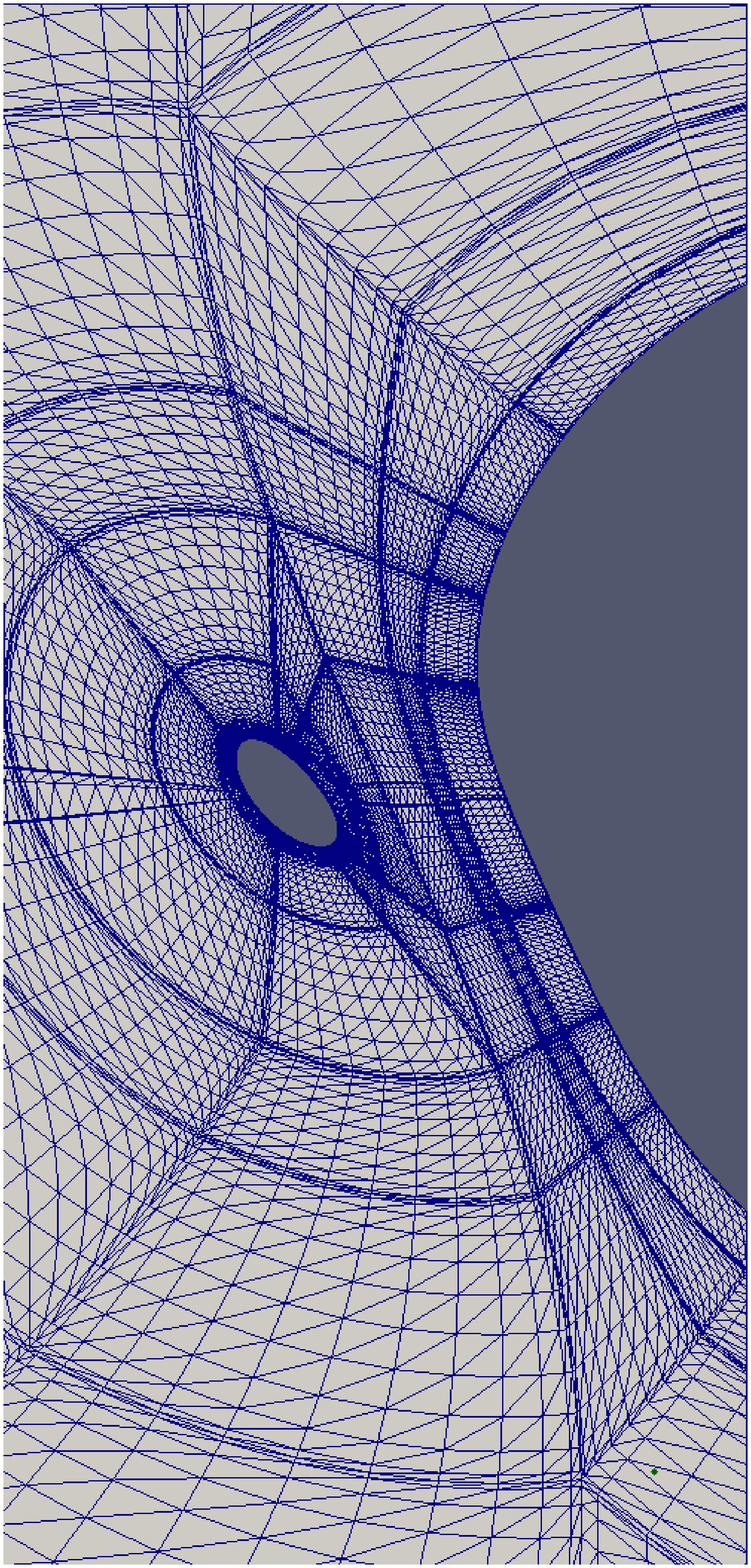}
  \end{flushleft}
  \end{minipage}
  \caption{
    Snapshot of the grid, viewed in the mapped frame,
    from a binary black hole evolution with $M_A=8M_B$
    shortly before the common horizon forms. Left: without
    \map{Skew}. Right: with \map{Skew}.
    The gray areas are excised regions.  In the left panel, the grid is
    compressed as the excision boundaries track increasingly skewed
    apparent horizons, and the evolution is terminated
    because of excessively
    large constraint violation in the compressed region.
    This problem is resolved by the inclusion of \map{Skew}.
  \label{fig:SimulationSkew}
  }
\end{figure}

The skew map, \map{Skew}, acts on the distorted-frame coordinates, 
in which the cutting plane is 
still perpendicular to the $x$-axis.
The skew map leaves the coordinates $y,z$
unchanged, but changes the $x$-coordinate in order to give a skewed
shape to the cutting plane, as shown in Fig.~\ref{fig:SimulationSkew}.
Let $x^i_C$ be
the intersection point of the line
segment connecting the excision boundary centers and the cutting plane.
The action of the skew map is defined
as
\begin{align}
  x^0 &\mapsto x^0 + W \sum_{j=1,2} \tan \left[ \Theta^j(t) \right] \cdot (x^j-x^j_C), \\
  x^j &\mapsto x^j, \;\;  \mbox{for} \;\; j=1,2
\end{align}
where $W$ is a radial Gaussian function centered around $x^i_C$, and
the angles $\Theta^j(t)$ are the time-dependent map parameters.
For $j=1,2$ the parameter $\Theta^j(t)$ 
is the angle between the mapped and unmapped $x^j$-axis when projected
into the $x^{3-j}=x^{3-j}_C$ plane. Note that the
line intersecting $x^i_C$ and parallel to the $x$-axis
is left invariant by the skew map\footnote{
The horizon centers $\distorted{\xi}^i_H$ are not invariant under the 
skew map because they do not necessarily lie on this line.
This deviation,
which is quantified in Eqs.~\eqref{eq:CenterAhx}--\eqref{eq:CenterAhz},
leads to
a coupling of $\mathcal{O}(Q)$ with the translation, rotation, and scaling maps.
}.
The width of the Gaussian $W$ is set such that $W$
is below machine precision at the innermost wave extraction sphere.
This implies that the spherical-shell subdomains used to evolve the metric
in the wave zone will not be affected by the skew map.  

The purpose of the skew map is to (as much as possible) align the
cutting plane with the surfaces of the apparent horizons in the region
where the surfaces are closest to the cutting plane.
We derive the control system in charge of setting the parameters
$\Theta^j$ by the following condition:  We demand that the angle
between the mapped cutting plane and the $x$-axis at $x^i_C$ be driven
to the (weighted) average of the angles at which the 
mapped
apparent horizons
intersect the same $x$-axis.

\begin{figure}
  \begin{center}
      \includegraphics[height=4cm]{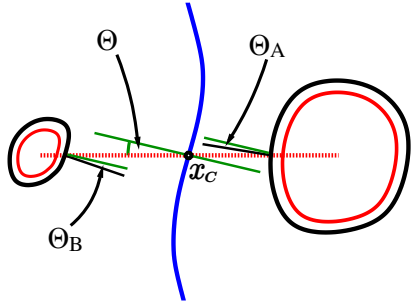}
  \end{center}
  \caption{A diagram of the skew map
   in the mapped frame.
   The horizon surfaces are drawn in black and the excision surfaces are drawn in red.
   The dotted
   red line represents the line
   segment connecting the excision surface centers,
   which is parallel to the $x$-axis.
   The normal to either horizon at the intersection point with this segment is
   represented by the straight black lines.
   The cutting plane is represented by the (skewed) blue curve,
   and the normal to this plane at $x^i_C$ is represented by the
   central green line.
     The green lines near the two excision surfaces are constructed
     parallel to the central green line, where parallelism is defined
     assuming a Euclidean background.
   There are three
   angles involved in the skew control system --
   the angle between the normal to the skewed cutting plane and the $x$-axis, $\Theta^j$,
   and the angle between the normal to either horizon and the
   normal to the skewed cutting plane, $\Theta^j_H$.
   Skew control acts to minimize $\Theta^j_H$.
   We measure $\Theta^j_H$ in the unmapped (distorted) frame, but $\Theta^j_H$
   is invariant under the skew map for $W\approx 1$.
  }
  \label{fig:SkewAngles}
\end{figure}

The input coordinates to the skew map are the coordinates in the
distorted frame.  
For each horizon, we therefore measure an angle in the distorted frame
as follows:
Let $x^{\rm Int}_H$ be the distorted-frame $x$-coordinate at which 
apparent horizon
$H$ intersects the line
segment connecting the centers of the excision boundaries.
For $j=1,2$, we calculate the normal to the surface at
this intersection point.
This normal
is projected  into the $x^{3-j}=x^{3-j}_C$ plane. 
We define $\Theta_H^j$ as the angle between the projected normal
and the $x$-axis.
Thus, projecting the
normal
into the $y={\rm const.}$ plane gives $\Theta_H^z$ and vice versa 
(see Fig.~\ref{fig:SkewAngles}).
We then compute a weighted average of the $\Theta^j_H$ such that the
horizon closer to the cutting plane has a larger effect on the skew angles,
\begin{equation}
\label{eq:SkewAverageAngle}
\Theta^j_{\rm Avg} = \frac{ w_A \Theta_A^j + w_B \Theta_B^j }
                         {w_A W(C_A) + w_B W(C_B)},
\end{equation}
where $w_A, w_B$ are averaging weights,
\begin{equation}\label{eq:SkewAveragingWeights}
  w_H = \exp\left[-\frac{x_C^0 - x^{\rm Int}_H}{x_C^0 - C^0_H}\right].
\end{equation}
Here $C^i_H$ is the center of the excision boundary $H$.

We measure $\Theta^j_{\rm Avg}$ in the distorted frame, and in this
frame the cutting plane is always normal to the $x$-axis.  Therefore,
thinking about the desired result in the distorted frame, we see that
the control system for the skew map should drive $\Theta^j_{\rm Avg}$
to zero.  This leads us to consider the following control error for
the skew angles: $Q^j_{\Theta}=\Theta^j_{\rm Avg}$.  Assuming that the function
$W$ is unity near
the apparent horizons, we find that
\begin{equation}
  \label{eq:SkewAverageAngleDerivative}
  \frac{\partial \Theta^j_{\rm Avg}}{\partial \Theta^j} = -1+{\cal O}(Q),
\end{equation}
in agreement with Eq.~(\ref{eq:Qgenericnonlinear}). In deriving
Eq.~(\ref{eq:SkewAverageAngleDerivative}), it is helpful to observe
that the partial
derivative in Eq.~(\ref{eq:SkewAverageAngleDerivative}) is taken
with the inertial-frame apparent horizon held fixed. 
For a fixed inertial-frame horizon, the only
map that can change the shape (as opposed to merely the center) 
of the distorted-frame
horizon (and thus $\Theta^j_{\rm Avg}$) is the skew map.

We do not use $Q^j_{\Theta} =\Theta^j_{\rm Avg}$ for the entire evolution,
however,
because at early times, when the coordinate distance between the
apparent horizons is larger than their combined radii,
the skew map is not needed. Furthermore, the skew map can cause difficulties
early during the run, especially
during the ``junk radiation'' phase when the horizons are oscillating in shape.
For this reason,
we gradually turn on the skew map as the black holes approach each other.
This is done by defining a roll-on function $g$ that is zero when the
horizons are far apart, and one when they are close together.
This roll-on function is defined as
\begin{equation}
  g = \frac{1}{2}\left[ 1-\tanh\left( 10\frac{x^{\rm Int}_A - x^{\rm Int}_B}
{C^0_A-C^0_B}-5 \right) \right].
\end{equation}
For values $g<10^{-3}$ the skew map is turned off
completely; this is not strictly necessary, but it saves some computation. 
Given the function $g$,
we define the control error as
\begin{equation}
\label{eq:SkewControlError}
Q^j_{\Theta}=g\Theta^j_{\rm Avg}-(1-g)\Theta^j.
\end{equation}
This control error drives the skew angles to zero when the black holes
are far apart and drives
$\Theta^j_{\rm Avg}$
to zero
as they
approach each other.


\subsection{Shape control}\label{sec:shape-control}

We define the shape map \map{Shape} as:
\begin{equation}
  x^i \mapsto x^i\left(1-\sum_H\frac{f_H(r_H,\theta_H,\phi_H)}{r_H}
  \sum_{\ell m} Y_{\ell m}(\theta_H,\phi_H)\lambda^H_{\ell m}(t)\right).
\label{eq:GridToShape}
\end{equation}
The index $H$ goes over each of the two excised regions $A$ and $B$, and
the map is applied to the grid-frame coordinates. The polar coordinates
$(r_H,\theta_H,\phi_H)$ centered about excised region $H$
are defined by Eqs.~\eqref{eq:PolarCoords1}--\eqref{eq:PolarCoords3},
the quantities $Y_{\ell m}(\theta_H,\phi_H)$ 
are spherical harmonics, and
$\lambda^H_{\ell m}(t)$ are expansion coefficients that parameterize the
map near excision region $H$.  The function $f_H(r_H,\theta_H,\phi_H)$ 
is chosen to be unity near excision region $H$ and zero near the
other excision region, so that the distortion maps for the two black holes 
are decoupled.  Specifically, 
$f_H(r_H,\theta_H,\phi_H)$ is determined as illustrated in 
Fig.~\ref{fig:CutSphere}.  For excision region $A$ in the figure, 
$f_A(r_A,\theta_A,\phi_A)=1$
between the excision boundary and the magenta surface,
it falls linearly to zero between the magenta and red surfaces,
and it is zero everywhere outside the red surface.
This means
that the gradient of $f_A(r_A,\theta_A,\phi_A)$ is discontinuous on the red
surface, the magenta surface, and the blue surfaces in the figure.  Because
we ensure that these discontinuities occur on subdomain boundaries, they cause
no difficulty with using spectral methods.
Around excision region $B$, $f_B(r_B,\theta_B,\phi_B)$ is chosen similarly.

In previous implementations of the shape map~\cite{Scheel2009}, the
functions $f_H(r_H,\theta_H,\phi_H)$ were chosen to be smooth
Gaussians centered around each excision boundary rather than to be
piecewise linear functions.  We find smooth Gaussians to be inferior
for two reasons.  The first is that piecewise linear functions are
easier and faster to invert (the inverse map is required for
interpolation to trial solutions during apparent horizon finding).
The second is that for smooth Gaussians, it is necessary to choose the
widths of the Gaussians sufficiently narrow so that the Gaussian for
excision region $A$ does not overlap the Gaussian for excision region
$B$ and vice versa, so that the maps and control systems for $A$ and
$B$ remain decoupled.  However, decreasing the width of the Gaussians
increases the Jacobians of the map, producing coordinates that are
stretched and squeezed nonuniformly.  We found that this form of
``grid-stretching'' significantly increased the computational
resources required to resolve the solution to a given level of
accuracy.  In other words, with smooth Gaussians we were forced to add
computational resources just to resolve the large Jacobians.

A map very similar to Eq.~\eqref{eq:GridToShape} is also described
in~\cite{Szilagyi:2009qz}. The difference compared to this work is the
choice of $f_H(r_H,\theta_H,\phi_H)$, which corresponds to a different
choice of domain decomposition. The control system used to choose the
map parameters $\lambda^H_{\ell m}(t)$ in~\cite{Szilagyi:2009qz} is
also different than what is described here.

We control the expansion coefficients $\lambda^H_{\ell m}(t)$ of
the shape map, Eq.~(\ref{eq:GridToShape}),
so that each excision boundary, as measured in the intermediate frame
$(t,\distorted{x}^i)$, has the same shape as the corresponding apparent horizon.
In other words, we desire
\begin{equation}
\frac{\distorted{r}_{\rm AH}}{\langle\distorted{r}_{\rm AH}\rangle} 
= \frac{\distorted{r}_{\rm EB}}{\langle\distorted{r}_{\rm EB}\rangle},
\label{eq:ShapeEquality}
\end{equation}
where for brevity we have dropped the index $H$ that labels the excision
boundary.  Here the angle brackets mean averaging over angles,
$\distorted{r}_{\rm AH}(\distorted{\theta},\distorted{\phi})$ 
is the radial coordinate of the 
apparent horizon defined in Eq.~\eqref{eq:StrahlkorperAH}, and
$\distorted{r}_{\rm EB}(\theta,\phi)$ is the radial coordinate of the
excision boundary, which from Eq.~\eqref{eq:GridToShape} can
be written
\begin{equation}
  \distorted{r}_{\rm EB} = r_{\rm EB} - 
  \sum_{\ell m} Y_{\ell m}(\distorted{\theta},\distorted{\phi})\lambda_{\ell m}(t).
\label{eq:GridToDistortedExcisionRadius}
\end{equation}
Here $r_{\rm EB}$ is the radius of the (spherical) excision boundary
in the grid frame.  In deriving Eq.~\eqref{eq:GridToDistortedExcisionRadius}
we have used the relations ${\cal M}_{\rm CutX}=\bm{1}$, 
$f(r,\theta,\phi)=1$, $\distorted{\theta}=\theta$,
and $\distorted{\phi}=\phi$, which hold on the excision boundary.

Combining Eqs.~\eqref{eq:StrahlkorperAH},~\eqref{eq:ShapeEquality},
and~\eqref{eq:GridToDistortedExcisionRadius} yields
\begin{equation}
\frac{r_{\rm EB} - \sum_{\ell m} Y_{\ell m}(\distorted{\theta},\distorted{\phi})
\lambda_{\ell m}(t)}%
{r_{\rm EB} - Y_{00} \lambda_{00}(t)} = 
\frac{\sum_{\ell m} \distorted{S}_{\ell m} 
Y_{\ell m}(\distorted{\theta}, \distorted{\phi})}%
{\distorted{S}_{00} Y_{00}},\label{eq:ShapesAreEqual}
\end{equation}
which we can satisfy by demanding that
\begin{equation}
  \lambda_{\ell m}(t)+
\distorted{S}_{\ell m} \frac{r_{\rm EB} - Y_{00} \lambda_{00}(t)}%
{\distorted{S}_{00} Y_{00}}=0, \quad \ell>0.
\end{equation}
Therefore, given an apparent horizon and given a value of $\lambda_{00}$,
a control system can be set up for each $(\ell,m)$ pair with $\ell>0$,
and the corresponding control errors are
\begin{equation}
Q_{\ell m} = -\lambda_{\ell m}(t)-
\distorted{S}_{\ell m} \frac{r_{\rm EB} - Y_{00} \lambda_{00}(t)}%
{\distorted{S}_{00} Y_{00}}, \quad \ell>0.\label{eq:Qlm}
\end{equation}
Driving these $Q_{\ell m}$ to zero produces an excision boundary that matches the
shape of the corresponding apparent horizon.

Equation~\eqref{eq:Qlm} determines $\lambda_{\ell m}(t)$ only for $\ell>0$,
and Eqs.~\eqref{eq:ShapeEquality}--\eqref{eq:Qlm} can be satisfied for
arbitrary values of the remaining undetermined map coefficient
$\lambda_{00}(t)$.  Determination of $\lambda_{00}(t)$ is complicated
enough that it is described in its own section,
Sec.~\ref{sec:size-control}.

Note that Eq.~(\ref{eq:Qlm}) does not satisfy
Eq.~(\ref{eq:Qdecoupled}) because $\partial Q_{\ell m}/\partial
\lambda_{00}=\distorted{S}_{\ell m}/\distorted{S}_{00}$, which
does not vanish even if all the control errors are zero.  For small
distortions, this coupling between $\lambda_{00}$ and $Q_{\ell m}$
seems to cause little difficulty.  However, at times when the shapes
and sizes of the horizons change rapidly (e.g.,  during the initial
``junk radiation'' phase, after the transition to a single excised
region when a common apparent horizon forms, and after rapid gauge
changes), simulations using Eq.~(\ref{eq:Qlm}) exhibit relatively
large and high-frequency oscillations in $Q_{\ell m}$ that usually
damp away but occasionally destabilize the evolution.  Construction of a
control system in which all $Q_{\ell m}$ are fully decoupled will be
addressed in a future work.

\subsection{CutX}
\label{sec:cutx}
\begin{figure}
\centerline{\includegraphics[scale=0.37]{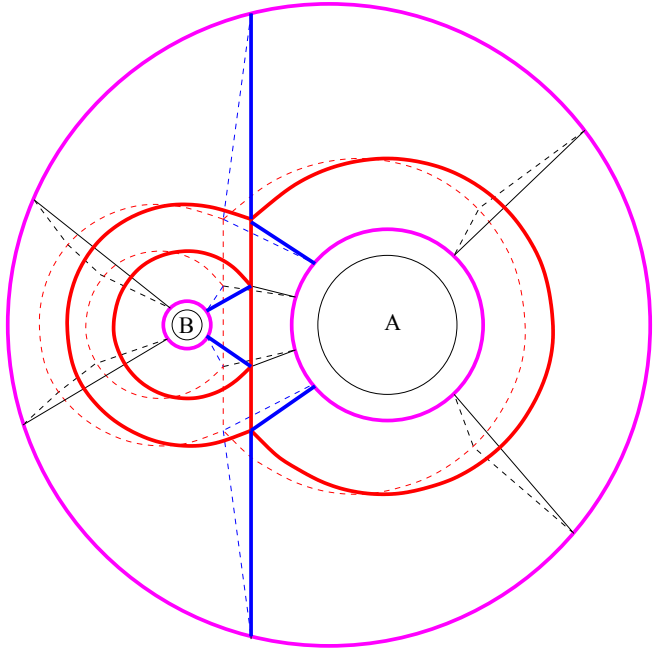}}
\caption{\label{fig:CutX} Two-dimensional projection of the
  domain decomposition near the two
  black holes A and B.  
  The function $\rho$ in \map{CutX} is zero on the magenta
  boundaries, one on the red boundaries, and goes linearly between zero and one
  in the ``radial'' direction between these boundaries.  
  The gradient of $\rho$ is discontinuous across the solid
  magenta, red, and blue boundaries. Dotted lines show the subdomain boundaries
  under the action of \map{CutX}. }
\end{figure}

The map \map{CutX} applies a translation along the 
grid-frame $x$-axis in the vicinity of the black holes, but without
moving the excision boundaries (or the surrounding spherical shells)
themselves.  The action of the map
is shown in Fig.~\ref{fig:CutX}.

The goal of this map is to allow for a slight motion of the cutting
plane toward the smaller excision boundary.  This is important for
binary black hole systems with mass ratio $q\gtrsim 8$.  As such a binary
gets closer to merger, the inertial-frame coordinate distance between
each excision boundary and the cutting plane decreases.
Eventually, this distance falls to zero for the larger excision boundary,
producing a coordinate
singularity in which the Jacobian of one of the other maps (often the
shape map) becomes infinite.  By pushing the cutting plane toward the smaller
excision boundary, the map \map{CutX} avoids this
singularity.  Even for evolutions in which the inertial-coordinate distance
between the larger excision boundary and the cutting plane remains
finite but becomes small, the map \map{CutX} prevents large
Jacobians from developing and thus increases numerical accuracy (because
it is no longer necessary to add computational resources to resolve the
large Jacobians).

\begin{figure}
  \begin{center}
      \includegraphics[height=4cm]{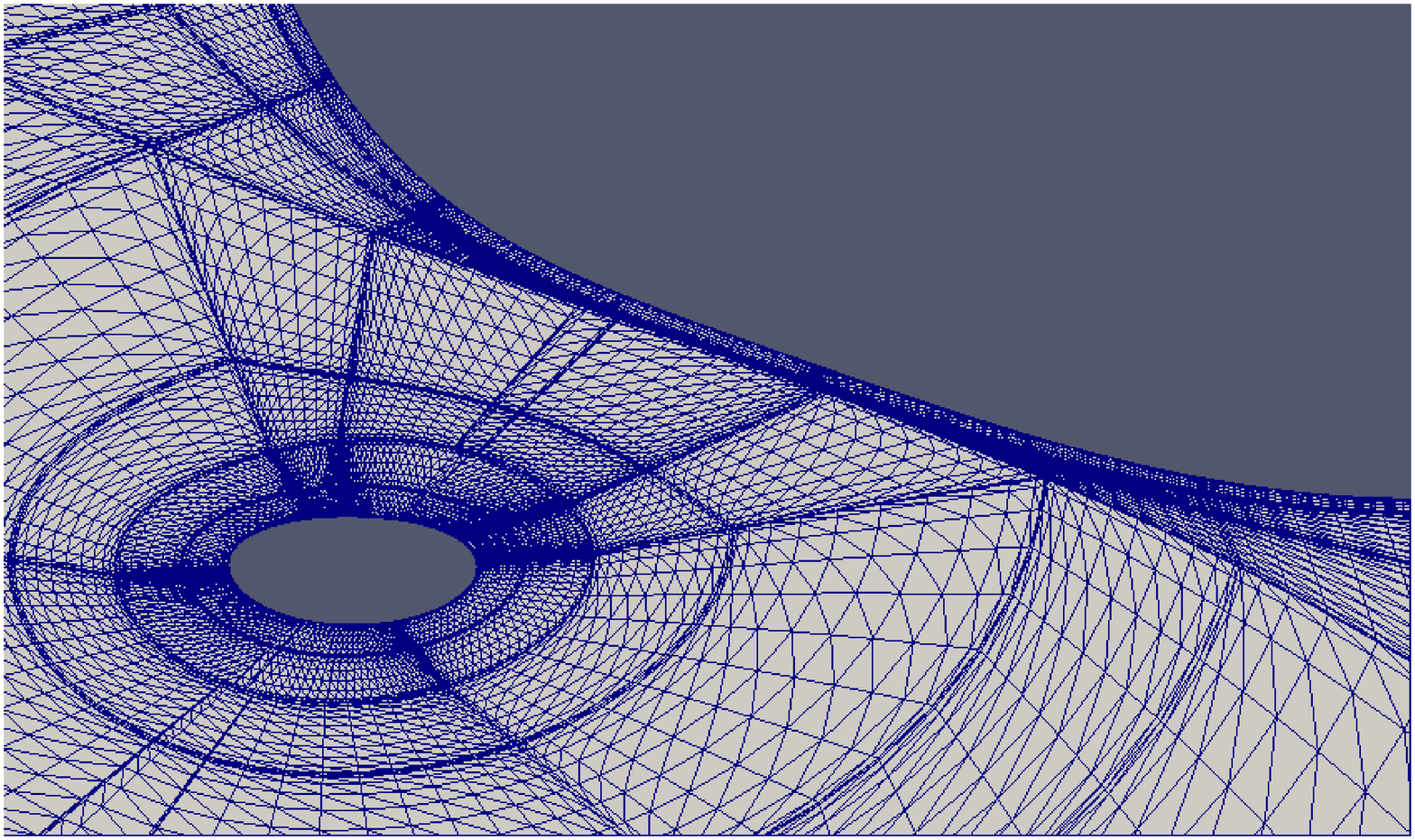}
  \end{center}
  \begin{center}
      \includegraphics[height=4cm]{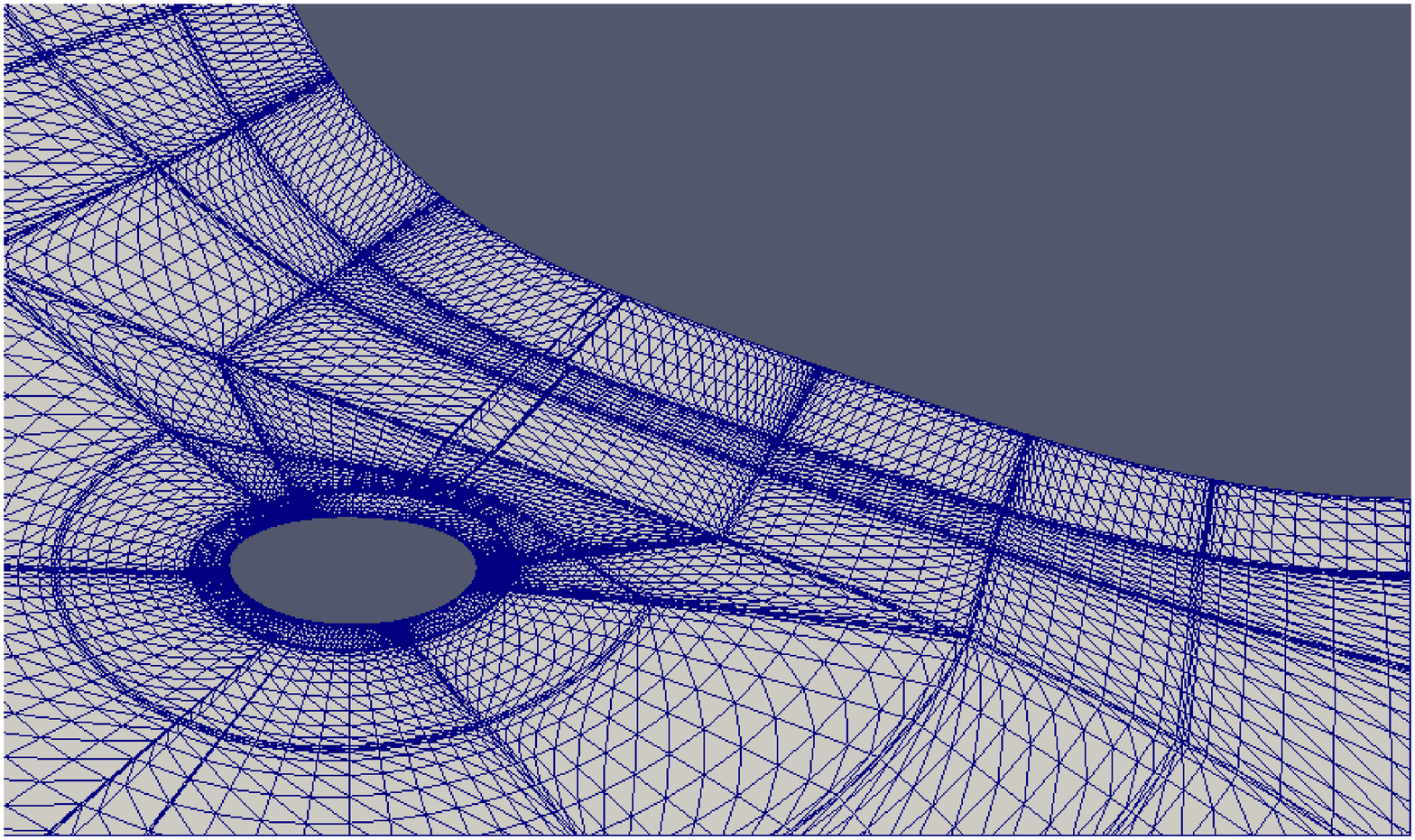}
  \end{center}
  \caption{
    Snapshot of the grid, viewed in the inertial frame, 
    from a binary black hole evolution with $M_A=8M_B$
    shortly before the common horizon forms.  Top: without
    \map{CutX}. Bottom: with \map{CutX}.
    The gray areas are excised regions.  In the top panel, the grid is
    compressed as the larger excision boundary approaches the cutting plane;
    this is especially evident in the long narrow subdomains immediately
    adjacent to the larger excision boundary.
    Using \map{CutX} relieves the grid compression.
    \label{fig:SimulationCutX}
  }
\end{figure}

Figure~\ref{fig:SimulationCutX} shows the domain decomposition in the
inertial frame for a binary black hole simulation with $q=8$.  The top
panel shows the case without \map{CutX}, and the compressed
grid near the larger excision boundary is evident.  The lower panel
shows the case with \map{CutX}, which removes the extreme
grid compression.
Looking at the Jacobian of the mapping from the inertial frame to
the grid frame shortly before merger, we find that the infinity norm
of the determinant of the Jacobian is twice as large in the case
without \map{CutX}.

The map \map{CutX} is written as
\begin{align}
\label{eq:CutX.x}
  x^0 &\mapsto x^0 + \rho(x^i) F(t), \\
  x^j &\mapsto x^j, \qquad j=1,2,
\end{align}
where $F(t)$ is adjusted by a control system.

We choose
$\rho(x^i)$
to be zero within either of the spherical
shell regions, i.e., inside the magenta spheres around $A$ and $B$ in
Fig.~\ref{fig:CutX}, and it is also zero outside the outer
magenta sphere in the same figure.  We set $\rho(x^i)=1$ on the
solid red boundaries in Fig.~\ref{fig:CutX}, and everywhere between
the two solid red boundaries that enclose excision boundary $B$.
Every other region on the grid is bounded by a smooth red boundary on
one side and a smooth magenta boundary on the other; in these regions
$\rho$ varies linearly between zero and one.  The solid blue
boundaries are locations (in addition to the solid red and solid
magenta boundaries) in which the gradient of $\rho$ is discontinuous.
Full details for the calculation of $\rho$ can be
found in \ref{sec:cutxweight}.

As with the skew map, the map \map{CutX} is inactive
for most of the inspiral.  Let
$x^{\rm Exc}_H$
be the distorted-frame $x$-coordinate of the intersection
of the excision boundary
$H$ with the
line segment
connecting the centers
of the excision boundaries. This is similar to $x^{\rm Int}_H$
defined earlier; the difference is that $x^{\rm Int}_H$ 
refers to
a point on
the apparent horizon
and $x^{\rm Exc}_H$ refers to
a point on
the excision boundary.
The quantity $x^{\rm Exc}_H$
is time-dependent because it depends on the shape
map.

The map \map{CutX} is turned off completely
as long as
\begin{equation}
  \frac{x^0_C-x^{\rm Exc}_H}{x^{\rm Exc}_H-C^0_H}\geq 
  \frac 12,
  \label{eq:CutXInactiveCondition}
\end{equation}
where $C^i_H$ are the excision boundary centers, and
$x^i_C$ are the coordinates of the intersection of the cutting plane
and the line segment connecting the centers of the excision boundaries, as
introduced in Sec.~\ref{sec:skew}.

Let $t_0$ be the coordinate time at which
the map \map{CutX} is activated.\footnote{
  Both \map{Skew} and \map{CutX} are turned on late in the run, but the
  condition triggering their activation is different,
  given the different nature of the problems they address.
  \map{Skew} is needed for all runs where the
  horizons eventually
  intersect the line segment connecting their excision centers
  at an angle sufficiently different from $\pi/2$.
  This will
  happen essentially for all runs except the simplest head-on
  collisions. \map{CutX}, on the other hand, is primarily for
  unequal-mass runs where the larger excision surface encroaches upon
  the cutting plane near merger.
}
We designate the target position $x=x_T$
of the cutting plane as
\begin{equation}
  x_{T} = \left. \frac{\Delta x_A \cdot x^{\rm Exc}_A
+ \Delta x_B \cdot x^{\rm Exc}_B}{\Delta x_A+ \Delta x_B } \right|_{t=t_0},
\end{equation}
where
\begin{equation}
  \Delta x_H = \left| x^{\rm Exc}_H-C^0_H \right|, \quad H=A,B.
  \label{eq:DeltaxH}
\end{equation}
Recall that $x^{\rm Exc}_H$ is time-dependent
(as it is measured in the distorted frame),
so we save the value of $x_{T}$ as calculated at the 
time when the map \map{CutX} is activated, rather than recalculating
$x_{T}$ at every measurement time.

Now that we have a target $x_{T}$, we could designate $x_{T}-x^0_C$ as
the target value to the function $F(t)$ by setting $Q_F=-F+x_{T}-x^0_C$
and have the control system drive $Q_F$ to zero, thus driving the
$x$-coordinate of the cutting plane to $x_{T}$.  However, because we
turn on the CutX control system suddenly at time $t_0$, we must be
more careful.  Turning on any control system suddenly will produce
some transient oscillations, unless the control error and its
relevant derivatives and integrals are all initially zero.  In the
case of the CutX map, which is turned on during a very dynamic part
of the simulation when excision boundaries need to be controlled very
tightly, these oscillations can
prematurely terminate
the run
by, for example,
pushing
an excision boundary outside its accompanying apparent horizon.

To turn on \map{CutX} gradually, we replace $x_T$ by a new time-dependent
target function $T(t)$ that
gradually approaches $x_{T}$ at late times but produces a control
error $Q_F$ with $Q_F=\partial Q_F/\partial t=0$ at the activation
time $t=t_0$.  We start by estimating the time $t^{\rm XC}_H$ at which
$x^{\rm Exc}_H$ will reach the cutting plane, where $H$ is either $A$ or $B$.
This is done by the method described in \ref{sec:estimatecrossing}.
We then let $t^{\rm XC}$ be the minimum of $t^{\rm XC}_A$ and $t^{\rm XC}_B$.
The smooth target function $T(t)$ is then defined by
\begin{align}
  T(t;t_0,x_T,t^{\rm XC}) = x_T &+
  \exp\left[-\left(\frac{t-t_0}{0.3 \cdot t^{\rm XC}}\right)^2\right]
  \nonumber
  \\
  \times 
  \bigg[
    x^0_C
    -x_{T}
    +F(t_0)
    &+
        \left.
  (t-t_0)
  \cdot
  \frac{\partial F(t) }{ \partial t}\right|_{t=t_0}
  \bigg] \, .
\end{align}
Designating $T-x^0_C$ as the target for $F(t)$
\begin{equation}
  Q_F = T-x^0_C-F
\end{equation}
leads to
\begin{equation}
  Q_F|_{t=t_0} = 0, \quad
  \left.\frac{\partial Q_F}{\partial t}\right|_{t=t_0} = 0,
\end{equation}
so at the activation time the control system does not produce
transients.  Furthermore, in the limit of small $Q_F$,
\begin{equation}
  x_C^0 + F(t)|_{t=t^{\rm XC}} \approx x_T,
\end{equation}
i.e., by the time the excision boundary would have touched
the cutting plane (and formed a grid singularity), the smooth target
function $T(t)$ has approached $x_T$, and the cutting plane
will have reached its designated target location, $x_T$.
As the run proceeds, the behavior of the cutting plane
is determined by the  map \map{CutX} while the motion
of the excision boundaries is determined by gauge dynamics
and the behavior of the other control systems. We
continue to monitor the distance between the cutting plane
and the excision boundaries, and if it is predicted to touch
within a time less than $\tau_d/0.15$, the \map{CutX} 
control system is reset, constructing a new target function
$T(t;t_0,x_T,t^{\rm XC})$, where $t_0$ is the time of the reset,
and $x_T,t^{\rm XC}$ are also recalculated based on the 
state of the grid at this reset time.

Each time a new target function $T$ is constructed, the
damping time $\tau_d$ of the \map{CutX} control system is set
to be $t^{\rm XC}/2$.

The control systems responsible for \map{CutX} and \map{Shape} are
decoupled, as \map{CutX} controls the location of the cutting
plane, leaving the excision boundary unchanged, while \map{Shape}
controls the shape of the excision boundary, leaving the location of
the cutting plane unchanged.  Recall that \map{CutX} and \map{Shape}
define the mapping from the grid frame to the distorted frame.  As the
apparent horizons are found in the distorted frame, and the other maps
only depend upon measurements of the horizons, \map{CutX} and
\map{Shape} are decoupled from the other maps.

\section{Size control}\label{sec:size-control}

In this section we discuss how we control the spherical part of the
map given by Eq.~\eqref{eq:GridToShape}, namely, the coefficients
$\lambda^H_{00}$ for each excision boundary $H$.  We apply the same
method to each excision boundary, so for clarity, in this section we
again drop the index $H$ from the coefficients $\lambda^H_{00}$ and
$S^H_{\ell m}$, and from the coordinates $(r_H,\theta_H,\phi_H)$.

\subsection{Characteristic speed control}

Controlling the size of the excision boundary is more complicated than
simply keeping the excision boundary inside the apparent horizon.
This is because black hole
excision requires conditions on the characteristic speeds of the system,
and if these conditions are not enforced they are likely to be violated.

The minimum characteristic speed at each excision boundary is given by
\begin{equation}
  v = -\alpha - \inertial{\beta}^i \inertial{n}_i - \inertial{n}_i
  \frac{\partial \inertial{x}^i}{\partial t},\label{eq:CharSpeedDef}
\end{equation}
where $\alpha$ is the lapse, $\inertial{\beta}^i$ is the shift, and
$\inertial{n}_i$ is the normal to the excision boundary pointing {\it out
  of the computational domain}, i.e., toward the center of the hole.
Here $(\inertial{t},\inertial{x}^i)$ are the inertial frame coordinates.  The
first two terms in Eq.~\eqref{eq:CharSpeedDef} describe the coordinate
speed of the ingoing (i.e., directed opposite to $\inertial{n}_i$) light
cone in the inertial frame, and the last term accounts for the motion
of the excision boundary (which is fixed in the grid frame) with
respect to the inertial frame.

In our simulations, we impose no boundary condition whatsoever at each
excision boundary.  Therefore, well-posedness requires that all of the
characteristic speeds, and in particular the minimum speed $v$, must be
non-negative; in other words, characteristics must flow into the hole.
In practice, if $v$ becomes negative, the
simulation is terminated, because a boundary condition is needed,
but we do not have one to impose.  
This can occur even when the excision boundary is
inside the horizon.  In this case, one might argue that if the 
simulation is able to continue without crashing (e.g. by becoming unstable
inside the horizon), that the solution outside 
the horizon would not be contaminated.  We have not explored this
possibility.  Instead, we choose to avoid this situation
by terminating the code if negative speeds are detected.

Therefore, we would like to control $\lambda_{00}$ in such a way 
that $v$ remains positive.  We start by writing $v$ in a way that
separates terms that explicitly depend on $\dot\lambda_{00}$ from
terms that do not.  To do this we expand the derivative in the
last term of
Eq.~\eqref{eq:CharSpeedDef} as
\begin{align}
  \frac{\partial \inertial{x}^i}{\partial t} &= 
  \frac{\partial \inertial{x}^i}{\partial \distorted{x}^a} 
  \frac{\partial \distorted{x}^a}{\partial t}
  \nonumber \\
  &=\frac{\partial \inertial{x}^i}{\partial \distorted{t}} 
    \frac{\partial \distorted{t}}{\partial t}+
    \frac{\partial \inertial{x}^i}{\partial \distorted{x}^j} 
    \frac{\partial \distorted{x}^j}{\partial t}
  \nonumber \\
  &=\frac{\partial \inertial{x}^i}{\partial \distorted{t}}+
    \frac{\partial \inertial{x}^i}{\partial \distorted{x}^j} 
    \frac{\partial \distorted{x}^j}{\partial t},
    \label{eq:CharSpeedPartialDerivExpansion}
\end{align}
where $a$ in the first line
of Eq.~(\ref{eq:CharSpeedPartialDerivExpansion}) is a four-dimensional spacetime
index, and the last line of Eq.~(\ref{eq:CharSpeedPartialDerivExpansion})
follows from $\partial\distorted{t}/\partial t = 1$. 
Inserting this into Eq.~(\ref{eq:CharSpeedDef}) yields 
\begin{equation}
 v = -\alpha - \inertial{\beta}^i \inertial{n}_i - \inertial{n}_i
 \frac{\partial \inertial{x}^i}{\partial \distorted{t}}
 - \distorted{n}_i \frac{\partial \distorted{x}^i}{\partial t},\label{eq:CharSpeedDef1}
\end{equation}
where $\distorted{x}^i$ is the frame that is obtained from the grid frame
by applying the distortion map; see Eq.~\eqref{eq:DistortionMapSymbol}.

We then use Eq.~\eqref{eq:GridToShape} to rewrite this as
\begin{align}
 v &= -\alpha - \inertial{\beta}^i \inertial{n}_i - \inertial{n}_i
 \frac{\partial \inertial{x}^i}{\partial \distorted{t}} \nonumber \\
 &+ \distorted{n}_i \frac{x^i}{r}
 \sum_{\ell m} Y_{\ell m}(\theta,\phi)\dot\lambda_{\ell m}(t),
 \label{eq:CharSpeedFormula}
\end{align}
where we have used the relations
$f(r,\theta,\phi)=1$ and ${\cal M}_{\rm CutX}=\bm{1}$ when evaluating
the distortion map \map{Distortion} on the excision boundary.

By combining all the terms that do not explicitly depend on
$\dot\lambda_{00}$ into a quantity $v_0$, we obtain  
\begin{equation}
   v = v_0 + 
\distorted{n}_i 
\frac{x^i}{r} Y_{00} \dot\lambda_{00}.
\end{equation}
Thus, the characteristic speed $v$
can be thought of as consisting of two parts: one part, $v_0$, that
depends on the position and shape of the excision boundary and the values of the
metric quantities there, and another part that depends on the average
speed of the excision boundary in the direction of the boundary normal.

We now construct a control system that drives the characteristic
speed $v$ to some target speed $v_T$. This is a control system
that controls the derivative quantity $\dot\lambda_{00}$, as opposed to 
directly controlling the map quantity $\lambda_{00}$.  We choose
\begin{equation}
Q = ({\rm min}(v)-v_T)/\langle-\Xi\rangle,\label{eq:CharSpeedControlQ}
\end{equation}
where
\begin{equation}
\Xi = \distorted{n}_i \frac{x^i}{r} Y_{00},
\end{equation}
the minimum is over the excision boundary,
and the angle brackets in Eq.~\eqref{eq:CharSpeedControlQ} refer to an
average over the excision boundary.
Note that $\Xi<0$ because $\distorted{n}_i$ points radially inward and
$x^i/r$ points radially outward; this means that $\dot Q = -\ddot\lambda_{00}$,
in accordance with our normalization choice for a control
system on $\dot\lambda_{00}$.

As in our other controlled map parameters, we demand that
$\dot\lambda_{00}$ is a function with a piecewise-constant second
derivative.  It is then easy to construct $\lambda_{00}$ as a function
with a piecewise-constant third derivative.

\begin{figure}[ht!]
  \begin{center}
      \includegraphics[height=5cm]{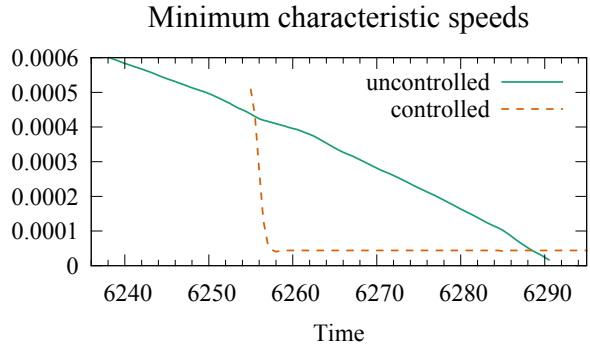}
  \end{center}
  \caption{Minimum characteristic speed on one of the excision
    surfaces of an equal mass binary with equal aligned spins of $\chi=0.97$,
    shown just before merger~\cite{Lovelace:2011nu}.  Two cases are
    shown: one without characteristic speed control (solid) and one
    that is restarted from the uncontrolled run at $t=6255M$ with
    characteristic speed control turned on and $v_T=5\times 10^{-5}$
    (dashed).  The uncontrolled run is terminated because of negative
    characteristic speeds shortly after
    $t=6290M$, but the controlled run continues through merger and
    ringdown.}
  \label{fig:CharSpeeds097}
\end{figure}

The control system given by Eq.~\eqref{eq:CharSpeedControlQ} with a
hand-chosen value of $v_T$ has been used successfully
\cite{Lovelace:2010ne,Lovelace:2011nu} in simulations of high-spin
binaries. Figure~\ref{fig:CharSpeeds097} illustrates why characteristic
speed control is crucial for the success of these simulations.

\subsection{Apparent horizon tracking}

The characteristic speed control described above has the disadvantage
that it requires a user-specified target value $v_T$.  If $v_T$ is
chosen to be too small, then small fluctuations (due to shape control,
the horizon finder, or simply numerical truncation error) can cause
the characteristic speed to become negative and spoil the simulation.

If $v_T$ is chosen to be too large, the simulation can also fail.  To
understand why, recall that characteristic speed control achieves the
target characteristic speed by moving the excision boundary and thus
changing the velocity term in Eq.~\eqref{eq:CharSpeedDef}.  So if
$v>v_T$ the control system moves the excision boundary radially
inward, and if $v<v_T$ the control system moves the excision boundary
radially outward.  If $v_T$ is too large, the control system can push
the excision boundary outward until it crosses the apparent horizon.
This halts the evolution because the apparent horizon can no longer be found.

One way to prevent the excision boundary from crossing the horizon
is to drive the excision boundary to some constant fraction of
the horizon radius, or in other words, drive the quantity $d/dt(\Delta r)$
to zero, where
\begin{equation}
\Delta r = 1 - \frac{\langle\distorted{r}_{\rm EB}\rangle}
                    {\langle\distorted{r}_{\rm AH}\rangle}
\end{equation}
is the relative difference between the average radius of the apparent
horizon (in the intermediate frame) and the average radius of the excision
boundary. Using Eqs.~\eqref{eq:GridToDistortedExcisionRadius}
and~\eqref{eq:StrahlkorperAH},
we can write
\begin{equation}
\frac{d}{dt}\Delta r = \frac{\dot\lambda_{00}}{\distorted{S}_{00}} +
                       \frac{\dot{\distorted{S}}_{00}}{\distorted{S}_{00}}
                       \left(1-\Delta r\right), 
\end{equation}
and therefore a control system that adjusts $\dot\lambda_{00}$ to achieve
$d/dt(\Delta r)=0$ can be obtained by defining
\begin{equation}
Q = \dot{\distorted{S}}_{00}(\Delta r-1)-\dot\lambda_{00}.\label{eq:State2Q}
\end{equation}

A slight generalization of this control system can be obtained by
demanding that $d/dt(\Delta r) = \dot{r}_{\rm drift}$, where 
$\dot{r}_{\rm drift}$ is
some chosen constant,
\begin{equation}
Q = \dot{\distorted{S}}_{00}(\Delta r-1)-\dot\lambda_{00}+\distorted{S}_{00}
\dot{r}_{\rm drift}.
\label{eq:State2DriftQ}
\end{equation}

We have not found the control systems defined by
Eqs.~\eqref{eq:State2Q} and~\eqref{eq:State2DriftQ} to be especially
useful on their own.  One drawback of these systems is that they do
not prevent the minimum characteristic speed $v$ from becoming
negative.  Instead, we use the horizon tracking control systems
described here as part of a more sophisticated control system
discussed in the next section.

\subsection{Adaptive switching of size control}
\label{sec-AdaptiveSizeControl}

Here we introduce a means of controlling $\lambda_{00}$ that combines
the best features of characteristic speed control and horizon tracking.
The idea is to continuously monitor the state of the system and switch
between different control systems as the evolution proceeds.

At fixed intervals during the simulation that we call ``measurement
times,'' we monitor the minimum characteristic speed $v$ and the
relative distance between the horizon and the excision boundary
$\Delta r$.  The goal of the control system is to ensure that both of
these values remain positive.  

Consider first the characteristic speed $v$.  Using the method
described in \ref{sec:estimatecrossing}, we determine whether
$v$ is in danger of becoming negative in the near future, and if so,
we estimate the timescale $\tau_v$ on which this will occur.
Similarly, we also determine whether $\Delta r$ will soon become
negative, and if so we estimate a corresponding timescale $\tau_{\Delta r}$.
Having estimated both $\tau_v$ and $\tau_{\Delta r}$, we
use these quantities to determine how to control $\dot\lambda_{00}$.
In particular, we switch between horizon tracking, Eq.~\eqref{eq:State2Q},
and characteristic speed control, Eq.~\eqref{eq:CharSpeedControlQ},
based on $\tau_v$ and $\tau_{\Delta r}$.

We do this by the following algorithm, which favors horizon tracking
over characteristic speed control unless the latter is
essential.  Assume that the control system for $\dot\lambda_{00}$ is
currently tracking the horizon, Eq.~\eqref{eq:State2Q}, and that
the current damping timescale is some value $\tau_d$.  If $v$ is in danger
of crossing zero according to the above estimate, and if $\tau_v<\tau_d$
and $\tau_v < \tau_{\Delta r}$, we then switch to characteristic speed
control, Eq.~\eqref{eq:CharSpeedControlQ}, we set $v_T=1.01 v$ where
$v$ is the current value of the characteristic speed (the factor 
$1.01$ prevents the algorithm from switching back from
characteristic speed control to horizon tracking on the very next
time step), and we reset the
damping time $\tau_d$ equal to $\tau_v$.  
Otherwise we continue to use Eq.~\eqref{eq:State2Q}, 
resetting $\tau_d=\tau_{\Delta r}$ if $\tau_{\Delta r}<\tau_d$.

Now assume that the control system for $\dot\lambda_{00}$ is
controlling the characteristic speed,
Eq.~\eqref{eq:CharSpeedControlQ}.  If $v$ is not
in danger of crossing zero, so that we no longer need active control
of the characteristic speed, then we switch to horizon tracking,
Eq.~\eqref{eq:State2Q}, without changing the damping time $\tau_d$.
If $v$ is in danger of crossing zero, and if
$\Delta r$ is either in no danger of crossing zero or if it will cross
zero sufficiently far in the future such that $\tau_{\Delta r} \ge
\sigma_1 \tau_d$, then we continue to use
Eq.~\eqref{eq:CharSpeedControlQ} with $\tau_d$ reset 
to ${\rm min}(\tau_d,\tau_v)$ 
so as to maintain control of the
characteristic speed.  Here $\sigma_1$ is a constant that we
typically choose to be about 20.  

The more complicated case occurs when
both $v$ and $\Delta r$ are
in danger of crossing zero and $\tau_{\Delta r} < \sigma_1 \tau_d$. In
this case,
we have two possible methods by which we attempt to prevent
$\Delta r$ from becoming negative.  We choose between these
two methods based on the values of $\tau_{\Delta r}$ and $\tau_d$,
and also by the value of a quantity 
that we call the \emph{comoving characteristic speed}:
\begin{align}
 v_c &= -\alpha - \inertial{\beta}^i \inertial{n}_i
        - \inertial{n}_i
        \frac{\partial \inertial{x}^i}{\partial \distorted{t}}
        \nonumber \\
 &+ \distorted{n}_i \frac{x^i}{r}
  \left[Y_{00}\dot{\distorted{S}}_{00}(\Delta r-1)+
  \sum_{\ell>0} Y_{\ell m}(\theta,\phi)\dot\lambda_{\ell m}(t)\right].
  \label{eq:ComovingCharSpeedFormula}
\end{align}
This quantity is the value that the characteristic speed $v$ would
have if horizon tracking (Eq.~(\ref{eq:State2Q})) 
were in effect and working perfectly.  
Equation~(\ref{eq:ComovingCharSpeedFormula}) is derived by assuming $Q=0$ in 
Eq.~(\ref{eq:State2Q}), solving for $\dot\lambda_{00}$, and substituting
this value into Eq.~(\ref{eq:CharSpeedFormula}).  The reason to consider
$v_c$ is that for ${\rm min}(v_c)<0$ 
(which can happen temporarily during a simulation), horizon tracking
is to be avoided, because horizon tracking will 
drive ${\rm min}(v)$ toward a negative value, namely ${\rm min}(v_c)$.

The first method of preventing $\Delta r$ from becoming negative
is to switch to
horizon tracking, Eq.~\eqref{eq:State2Q}; we do this if ${\rm min}(v_c)>0$
and if $\tau_{\Delta
  r}<\sigma_2 \tau_d$, where $\sigma_2$ is a constant that
we typically choose to be 5.  The second method, which we employ if
$\sigma_2\tau_d < \tau_{\Delta r} < \sigma_1 \tau_d$
or if ${\rm min}(v_c)<0$, is to continue
to use characteristic speed control, Eq.~\eqref{eq:CharSpeedControlQ},
but to reduce the target characteristic speed $v_T$ by multiplying
its current value by some fraction
$\eta$ (typically $0.25$).  By reducing the
target speed $v_T$, we reduce the outward speed of the excision
boundary, and this increases $\tau_{\Delta r}$.  The reason
we use the latter method for ${\rm min}(v_c)<0$ is that the
former method, horizon tracking, will drive $v$ toward $v_c$ and we
wish to keep $v$ positive. The reason we use the
latter method for $\sigma_2\tau_d < \tau_{\Delta r} < \sigma_1 \tau_d$
is that in this range of timescales the control system is already
working hard to keep $v$ positive, therefore a switch to
horizon tracking is usually followed by a switch back to 
characteristic speed control in only a few time steps,
and rapid switches between different types of control
make it more difficult for the
control system to remain locked.  The constants $\eta$, $\sigma_1$,
and $\sigma_2$ are adjustable, and although the details of the
algorithm (i.e., which particular quantity is controlled at which
particular time) are sensitive to the choices of these constants, the
overall behavior of the algorithm is not, provided $1 < \sigma_2 <
\sigma_1$.

The overall behavior of this algorithm is to cause $\dot \lambda_{00}$
to settle to a state in which both $v>0$ and $\Delta r>0$ for a long
stretch of the simulation.  Occasionally, when either $v$ or $\Delta
r$ threatens to cross zero as a result of evolution of the
metric or gauge, the algorithm switches between characteristic
speed control and horizon tracking several times and then settles into
a new near-equilibrium state in which both $v>0$ and $\Delta r>0$.

The actual value of $\Delta r$ and of $v$ in the near-equilibrium
state is unknown a priori, and in particular this algorithm can in
principle settle to values of $\Delta r$ that are either small enough
that control system timescales must be very short in order to prevent
$\Delta r$ from becoming negative, or large enough that excessive
computational resources are needed to resolve the metric quantities
deep inside the horizon.  To prevent either of these cases from occurring,
we introduce a constant correction velocity $\dot{r}_{\rm corr}$,
a nominal target value $\Delta r_{\rm target}$, and a nominal target
characteristic speed $v_{\rm target}$. We currently
choose these to be $\dot{r}_{\rm corr}=0.005$, 
$\Delta r_{\rm target}=0.08$, and $v_{\rm target}=0.08$. 
These quantities are used only
to decide whether to replace Eq.~\eqref{eq:State2Q} by
Eq.~\eqref{eq:State2DriftQ} when the algorithm chooses
horizon tracking, as
described below; these
quantities are not used when the algorithm chooses
characteristic speed control.

First we consider the case where $\Delta r$ is
too large during horizon tracking. If
$\Delta r>1.1\Delta r_{\rm target}$, then we replace Eq.~\eqref{eq:State2Q} with
Eq.~\eqref{eq:State2DriftQ}, and we use $\dot{r}_{\rm drift}=-\dot{r}_{\rm corr}$.
We continue to use $\dot{r}_{\rm drift}=-\dot{r}_{\rm corr}$ until 
$\Delta r<\Delta r_{\rm target}$, at which point we set $\dot{r}_{\rm drift}=0$.

The case where $\Delta r$ is too small during horizon tracking
is more complicated because we want to be careful so as to
not make the characteristic speeds decrease, since it is more difficult
to recover from decreasing characteristic speeds when $\Delta r$ is small. 
In this case, if $\Delta r<0.9\Delta r_{\rm target}$,
$v<0.9v_{\rm target}$, and $\langle n^k\partial_k v_c\rangle>0$,
then we replace Eq.~\eqref{eq:State2Q} with
Eq.~\eqref{eq:State2DriftQ}, and we use $\dot{r}_{\rm drift}=v_{\rm corr}$.
Here $v_c$ is the comoving characteristic speed defined in 
Eq.~(\ref{eq:ComovingCharSpeedFormula}) and the angle brackets refer
to an average
over the excision boundary.
The condition on the gradient of $v_c$ ensures that $v_c$ will increase if 
$\Delta r$ is increased; otherwise a positive $\dot{r}_{\rm drift}$ will threaten
the positivity of the characteristic speeds.  We continue to use
$\dot{r}_{\rm drift}=v_{\rm corr}$ until either
$v>0.9v_{\rm target}$, $\Delta r>0.9\Delta r_{\rm target}$, 
$\langle n^k\partial_k v_c\rangle<0$, or if 
$v$ will cross $v_{\rm target}$ or $\Delta r$ will 
cross $\Delta r_{\rm target}$ within a time less than the current
damping time $\tau_d$
(where crossing times are determined by the procedure
of \ref{sec:estimatecrossing}).
When any of these conditions occur, we
switch to $\dot{r}_{\rm drift}=0$, and we prohibit switching back
to a positive $\dot{r}_{\rm drift}$ until either 
$v>0.99v_{\rm target}$ or $\Delta r>0.99\Delta r_{\rm target}$, or until
both $v$ and $\Delta r$ stop increasing in time.  These last conditions
prevent the algorithm from oscillating rapidly between
positive and zero values of $\dot{r}_{\rm drift}$.

\section{Merger and ringdown}
\label{sec:merger-ringdown}

The maps and control systems in Sec.~\ref{sec:control-systems-maps}
were discussed in the context of a domain decomposition with two
excision boundaries, one inside each apparent horizon.  The skew and
CutX maps were introduced specifically to handle difficulties that
occur when two distorted excision boundaries approach each other.  At
some point in the evolution as the two black holes become sufficiently
close, a common apparent horizon forms around them.  After this
occurs, the simulation can be simplified considerably by constructing
a new domain decomposition with a single excision boundary placed just
inside the common horizon.  The evolved variables are interpolated
from the old to the new domain decomposition, and the simulation then
proceeds on the new one.

The algorithm for transitioning to a new grid with a single excision
boundary is fundamentally the same as that described
in~\cite{Scheel2009,Szilagyi:2009qz}, but there have been improvements
in the maps and the control systems, so for completeness we describe
the procedure here.

At some time $t=t_m$ shortly after a common horizon forms, 
we define a new, post-merger set of grid coordinates
$\newgrid{x}^i$ and a corresponding new domain decomposition composed only of
spherical shells.  In $\newgrid{x}^i$ coordinates, the new excision
boundary is a sphere centered at the origin.  We also define a post-merger map
$\inertial{x}^i = \map{Ringdown} \newgrid{x}^i$, where
\begin{equation}
  \label{eq:MapSequenceRingdown}
\begin{array}{ll}
\map{Ringdown} =&
\map{TranslationRD}
\circ\map{Rotation}
\circ\map{ScalingRD}
\circ\map{ShapeRD}.
\end{array}
\end{equation}
Note that the post-merger grid coordinates differ from the pre-merger
grid coordinates, but there is only one set of inertial coordinates
$\inertial{x}^i$.  Recall that in the dual-frame
picture~\cite{Scheel2006}, the inertial-frame components of tensors
are stored and evolved, so the evolved variables are continuous at
$t=t_m$ even though the grid coordinates are not.

The post-merger translation map $\map{TranslationRD}$ and its control system
are the same as discussed in Sec.~\ref{sec:translation}, except
that $\map{TranslationRD}$ translates with respect to the origin of
the $\newgrid{x}^i$ coordinate system, which is different from the
origin of the $\grid{x}^i$ coordinate system.  The post-merger translation
control system keeps the common apparent horizon centered at the
origin in the post-merger grid frame.  Note that the center of the common
apparent horizon does not necessarily lie on the same line as the
centers of the individual apparent horizons
(see~\cite{Lovelace:2009} for an example).  This means that
$\map{TranslationRD}$ is not continuous with the pre-merger translation map
$\map{Translation}$, except near the outer boundary where both maps
are the identity.

To evolve a distorted single black hole we do not need a rotation map.  However,
if rotation is turned off suddenly at $t=t_m$, then the outer boundary
condition (which is imposed in inertial coordinates) is not smooth in
time, and we see a pulse of gauge and constraint violating modes
propagate inward from the outer boundary because of this sharp change
in the boundary condition.  Therefore, we use the same rotation map
$\map{Rotation}$ before and after merger, and we slow down the
rotation gradually. To accomplish this, instead of adjusting the map
parameters $(\vartheta, \varphi)$ by a control system, for $t\ge t_m$ 
we set these parameters
equal to prescribed functions that approach a constant at late
times. We set
\begin{equation}
  \vartheta(t) = A + \left[B+C(t-t_m)\right]e^{-(t-t_m)/\tau_{\rm roll}},
\label{eq:RolloffTheta}
\end{equation}
where the constants $A$, $B$, and $C$ are determined by demanding that
$\vartheta(t)$ and its first two time derivatives are continuous at
$t=t_m$. The parameter $\varphi$ obeys an equation of the same form.
We choose $\tau_{\rm roll}=20M$.

The post-merger scaling map $\map{ScalingRD}$ is written
\begin{equation}
  \newgrid{R} \mapsto \left[1+\frac{b(t)-b(t_m)}{b(t_m)}\sin^2\left(\frac{\pi \newgrid{R}}
  {2 \newgrid{R}_{\rm OB}}\right)\right]\newgrid{R},
  \label{eq:ExpansionMap2}
\end{equation}
where $\newgrid{R}$ is the post-merger grid-coordinate radius, and
$\newgrid{R}_{\rm OB}$ is the outer boundary radius in the post-merger
grid coordinates.  The function $b(t)$ is given by Eq.~(\ref{eq:ScalingMapb})
before and after merger.
We set $\newgrid{R}_{\rm OB}=b(t_m)R_{\rm OB}$ so
that at $t=t_m$, $\map{ScalingRD}$ matches $\map{Scaling}$ at the
outer boundary.  Note that the post-merger scaling map is the identity near
the merged black hole; the only purpose of this map is
to keep the inertial-coordinate location of the outer boundary smooth
in time at $t=t_m$. 

The post-merger shape map $\map{ShapeRD}$ and its control system are the same
as discussed in Secs.~\ref{sec:shape-control}
and~\ref{sec:size-control}, except the map is centered about the
origin of the $\newgrid{x}^i$ coordinate system, the sum over excised
regions in Eq.~(\ref{eq:GridToShape}) runs over only one region (which
we call excision region C), and the function
$f_C(r_C,\theta_C,\phi_C)$ appearing in Eq.~(\ref{eq:GridToShape}) is
\begin{equation}
  f_C(r_C,\theta_C,\phi_C) = \left\{
    \begin{array}{ll}
     (r_C-r_{\rm max})/(r_{\rm EB}-r_{\rm max}),&r_{\rm EB}\le r_C<r_{\rm max}\\
      0,&r_C \ge r_{\rm max}\\
    \end{array}
    \right.
\end{equation}
where $r_{\rm max}$ is the radius of a (spherical) 
subdomain boundary that is well
inside the wave zone.  We typically choose $r_{\rm max} = 32 r_{\rm EB}$.

Once the post-merger map parameters are initialized, as discussed
below, we interpolate all variables from the pre-merger grid to the 
post-merger grid.
The inertial coordinates $\inertial{x}^i$ are the same before and
after merger, so this interpolation is easily accomplished using
both the pre-merger and post-merger maps, e.g.
$F(x^i)=F(\mathcal{M}^{-1}\map{Ringdown} \newgrid{x}^i)$ for some
function $F$.  After interpolating all variables onto the post-merger grid, we
continue the evolution.  This entire process --- from detecting a common
apparent horizon to continuing the evolution on the new domain --- is
done automatically.

\subsection{Initialization}
\label{sec:initialization}

All that remains for a full specification of \map{Ringdown} is to
describe how parameters of the two discontinuous maps $\map{ShapeRD}$ and 
$\map{TranslationRD}$ are initialized at $t=t_m$.  To do this,
we first construct a temporary distorted
frame with coordinates $\newdistorted{x}^i$, defined by
\begin{equation}
  \inertial{x}^i = \map{Translation}
                   \circ\map{Rotation}
                   \circ\map{ScalingRD} 
                   \newdistorted{x}^i.
\label{eq:TempDistortedCoords}
\end{equation}
The coordinates $\newdistorted{x}^i$ are the same as the post-merger
distorted coordinates except that the pre-merger translation map
$\map{Translation}$ is used in Eq.~(\ref{eq:TempDistortedCoords}).
We then represent the common apparent horizon (which is already
known in $\inertial{x}^i$ coordinates) in $\newdistorted{x}^i$
coordinates:
\begin{equation}
  \newdistorted{x}^i_{\rm AH}(\newdistorted{\theta}_C,\newdistorted{\phi}_C,t) 
= \newdistorted{x}^i_{\rm AH0}(t)
  + \newdistorted{n}^i \sum_{\ell m} 
\newdistorted{S}^C_{\ell m}(t) Y_{\ell m}(\newdistorted{\theta}_C, 
\newdistorted{\phi}_C).
\label{eq:StrahlkorperAHC}
\end{equation}
The expansion coefficients $\newdistorted{S}^C_{\ell m}(t)$ and
the horizon expansion center $\newdistorted{x}^i_{\rm AH0}(t)$ are computed
using the (inertial frame) common horizon surface plus
the (time-dependent) map defined in 
Eq.~(\ref{eq:TempDistortedCoords}).  
Here $(\newdistorted{\theta}_C,\newdistorted{\phi}_C)$ are
polar coordinates centered about $\newdistorted{x}^i_{\rm AH0}$, and
$\newdistorted{n}^i$ is a unit vector corresponding to the direction
$(\newdistorted{\theta}_C,\newdistorted{\phi}_C)$.  
Equation~(\ref{eq:StrahlkorperAHC}) is the
same expansion as Eq.~(\ref{eq:StrahlkorperAH}), except here we write
the expansion to explicitly include the center $\newdistorted{x}^i_{\rm AH0}(t)$.
In Eq.~(\ref{eq:StrahlkorperAHC}) we write
$\newdistorted{S}^C_{\ell m}(t)$ and $\newdistorted{x}^i_{\rm AH0}(t)$
as functions of time because we compute them at several discrete times
surrounding the matching time $t_m$.  By finite-differencing in time,
we then compute first and second time derivatives 
of $\newdistorted{S}^C_{\ell m}$ and $\newdistorted{x}^i_{\rm AH0}$ at $t=t_m$.

Once we have $\newdistorted{S}^C_{\ell m}$ and its time derivatives,
we initialize $\lambda^C_{\ell m}(t_m) = -\newdistorted{S}^C_{\ell m}(t_m)$,
where $\lambda^C_{\ell m}(t)$ are the map parameters appearing 
in Eq.~(\ref{eq:GridToShape}), and the minus sign accounts for the sign
difference between the definitions of Eqs.~(\ref{eq:GridToShape}) 
and~(\ref{eq:StrahlkorperAHC}). We do this for
all $(\ell,m)$ except for $\ell =m=0$: we set $\lambda^C_{00}(t_m)=0$, 
thus defining the radius of the common apparent horizon in the post-merger
grid frame.  For all $(\ell,m)$ including $\ell =m=0$ we set
$d\lambda^C_{\ell m}/dt\left.\right|{}_{t=t_m} 
= -d\newdistorted{S}^C_{\ell m}/dt\left.\right|{}_{t=t_m}$,
and similarly for the second derivatives.

Note that the temporary distorted coordinates $\newdistorted{x}^i$ are
incompatible with the assumptions of our control system, because the
center of the excision boundary $\newdistorted{x}^i_{\rm AH0}(t)$ in
these coordinates is time-dependent.  We therefore define new
post-merger distorted coordinates $\distorted{x}^i$ by
\begin{equation}
  \inertial{x}^i = \map{TranslationRD}
                   \circ\map{Rotation}
                   \circ\map{ScalingRD} 
                   \distorted{x}^i,
\label{eq:NewDistortedCoords}
\end{equation}
where
\begin{equation}
  \distorted{x}^i = \newdistorted{x}^i-\newdistorted{x}^i_{\rm AH0}(t).
  \label{eq:NewDistortedCoords2}
\end{equation}
If we denote the map parameters in $\map{Translation}$ by $T_0^i$, the
map parameters in $\map{TranslationRD}$ by $T^i$,
and if we denote $\map{Rotation}\circ\map{ScalingRD}$ by the matrix
$M^i_j$, then Eqs.~(\ref{eq:NewDistortedCoords}) 
and~(\ref{eq:NewDistortedCoords2}) require
\begin{equation}
  T^i(t) = T_0^i(t) + M^i_j(t)\newdistorted{x}^j_{\rm AH0}(t).
  \label{eq:PostMergerTranslationMap}
\end{equation}
Here we have assumed that $f(R)$ appearing in
Eq.~(\ref{eq:TranslationMap}) is unity in the vicinity of the horizon,
so we can treat the translation map as a rigid translation near
the horizon.
We initialize $T^i$ and its first two time derivatives at $t=t_m$
according to Eq.~(\ref{eq:PostMergerTranslationMap}).  Note that the
distorted-frame horizon expansion coefficients
$\newdistorted{S}^C_{\ell m}$ and hence the initialization of
$\lambda^C_{\ell m}(t)$ are unchanged by the change in translation map
because $\distorted{x}^i$ and $\newdistorted{x}^i$ differ only by an
overall translation, Eq.~(\ref{eq:NewDistortedCoords2}), which leaves
angles invariant.

\section{Control systems for efficiency}
\label{sec:non-map-systems}

Although the most important use of control systems in SpEC is to
adjust parameters of the mapping between the inertial and grid
coordinates, another situation for which control systems are helpful
is the approximation of functions that vary slowly in time, are needed
frequently during the simulation, but are expensive to compute.

For example, the average radius of the apparent horizon, $r_{\rm AH}$,
is used in the control system for $\dot \lambda_{00}$, which is
evaluated at every time step.  Evaluation at each time step is
necessary to allow the control system for $\dot \lambda_{00}$ to
respond rapidly to sudden changes in the characteristic speed or the
size of the excision boundary, as may occur after regridding, after
mesh refinement changes, or when other control systems (such as size
control) are temporarily out of lock.

However, computing the apparent horizon (and thus its average radius
$r_{\rm AH}$) at every time step is prohibitively expensive.  To significantly
reduce the expense, 
we define a function with piecewise-constant second derivative,
$r^{\rm appx}_{\rm AH}(t)$, we define a control error
\begin{equation}
Q = r_{\rm AH}-r^{\rm appx}_{\rm AH},
\end{equation}
and we define a control system that drives $Q$ to zero.  We then pass
$r^{\rm appx}_{\rm AH}(t)$ instead of $r_{\rm AH}$ to those functions
that require the average apparent horizon radius at each time step.
The control error $Q$, and thus the expensive computation of the apparent
horizon, needs to be evaluated only infrequently, i.e., on the timescale
on which the average horizon radius is changing, which may be
tens or hundreds of time steps.

\section{Summary}
\label{sec:summary}

In simulations of binary black holes, we use a set of time-dependent
coordinate mappings to connect the asymptotically inertial frame (in
which the black holes inspiral about one another, merge, and finally
ringdown) to the grid frame in which the excision surfaces
are stationary and spherical. The maps are described
by parameters that are adjusted by a control system to follow the
motion of the black holes, to keep the excision surface just inside the
apparent horizons of the holes, and also to prevent grid compression.
We take care to decouple the control systems and choose stable control
timescales.

The scaling, rotation, and translation maps are used to track the
overall motion of the black holes in the inertial frame.  These are
the most important maps during the inspiral phase of the evolution, as
the shapes of the horizons remain fairly constant after the relaxation
of the initial
data.  As the binary approaches merger, the horizon shapes
begin to distort, and shape and size control start to become important.
Shape and size control are especially
crucial for unequal-mass binaries and for black holes with
near-extremal spins. In the latter case,
the excision surface (which must remain an
outflow boundary with respect to the characteristics of the evolution
system) needs to be quite close to the horizon.

In order to decouple the shape maps of the individual black holes, our
grid is split by a cutting plane at which the shape maps reduce to the
identity map.  As the black holes merge, a skew map is introduced in
order to align the cutting plane with the excision surfaces (see
Fig.~\ref{fig:SimulationSkew}) and to minimize grid compression between
the two black holes.

Finally, the implementation of CutX control was necessary to complete
the merger of high mass-ratio configurations.  In these systems, 
the excision boundaries approach the cutting plane asymmetrically as
the black holes merge, thus compressing the grid between the cutting
plane and the nearest excision boundary.
CutX control translates the cutting plane to keep it centered between the two
excision boundaries, alleviating this grid compression.

We find that we need all of the maps described in this paper with a
sufficiently tight control system in order to robustly simulate a wide
range of the parameter space of binary black hole
systems~\cite{Szilagyi:2009qz,Lovelace:2010ne,Buchman:2012dw}.  Many
of these maps and control systems are also used in our simulations of
black hole -- neutron star
binaries~\cite{Duez:2008rb,Foucart:2010eq,FoucartEtAl:2011}. 

\appendix

\section{Implementation of exponentially-weighted averaging}
\label{sec:ExpAverager}

Our exponentially-weighted averaging scheme uses an averaging timescale
$\tau_{\rm avg}$ to smooth noisy quantities $F$ that are measured at
intervals $\tau_m := t_k - t_{k-1}$, where $F$ is the measured value of
the control error $Q$, its integral, or its derivatives.
Recall from Sec.~\ref{sec:DynamicTimescaleAdjustment} that we typically choose
 $\tau_{\rm avg}\sim 0.25\tau_d$ and
$\tau_m \sim 0.075\min(\tau_d)$.

The averaging is implemented by solving the following system of
ordinary differential equations (ODEs):
\begin{align}
\frac{d}{dt}W &= 1-\frac{W}{\tau\sub{avg}},\\
\frac{d}{dt}(W \tau) &= t-\frac{W\tau}{\tau\sub{avg}},\\
\frac{d}{dt}(W F_{\rm avg}) &= F-\frac{WF_{\rm avg}}{\tau\sub{avg}},
\end{align}
where the evolved variables are a weight factor $W$,
the average value $F_{\rm avg}$, and the effective time $\tau$ at which $F_{\rm
  avg}$ is calculated.  The ODEs are solved approximately using
backward Euler differencing,
which results in the recursive equations
\begin{align}
W_k &= \frac{1}{D}\left(\tau_m + W_{k-1}\right),         \\
\tau_k &= \frac{1}{DW_k}\left( \tau_m t_k + W_{k-1} \tau_{k-1}\right),    \\
F_{\rm avg}^k &= \frac{1}{DW_k}\left[\tau_m F(t_k) + W_{k-1}F_{\rm avg}^{k-1}\right],\label{eq:ThisF}
\end{align}
where $D=1+\tau_m/\tau\sub{avg}$.
The recursion is initialized with
$W_0=0$, $\tau_0=t_0$, and $F_{\rm avg}^0=F(t_0)$.

In the control law equations,
Eqs.~(\ref{eq:pid}) and~(\ref{eq:pnd}),
we want to use the averaged value at $t_k$ instead of $\tau_k$.
Therefore, to adjust for the offset induced by averaging,
$\delta t_k = t_k - \tau_k$, we evaluate at $t_k$
an approximate Taylor series for
$F_{\rm avg}^k$ expanded about $\tau_k$
\begin{equation}
F_{\rm avg}(t_k) = \sum_{m=0}^{N-n} \frac{\delta t^m}{m!}\left(\frac{d^mF}{dt^m}\right)_{\rm avg}^k,
\label{eq:FixOffset}
\end{equation}
where $n$ is defined by $F := d^nQ/dt^n$
(for convenience, $F$ represents the integral of $Q$ when $n=-1$),
and $N$ is
the same as in Eq.~\eqref{eq:GenericMapParam}, where it is defined as
the number of derivatives used to represent the map parameter.
Note that this approximation matches the true Taylor series
inasmuch as $W$ is constant.

For the highest order derivative of $Q$, where $n=N$, we can no longer
directly adjust for the offset
because Eq.~\eqref{eq:FixOffset} reduces to
\begin{equation}
F_{\rm avg}(t_k) = F_{\rm avg}^k.
\end{equation}
We would need to take additional derivatives of $Q$ to
provide a meaningful correction, but this is exactly what we want to avoid
because of the noisiness of derivatives.
Therefore, to circumvent taking higher order derivatives, we
substitute $F_{\rm avg}(t_k)$ into the control law, e.g. 
Eq.~\eqref{eq:pidode} for PID, and then solve for $d^NQ/dt^N$.

\section{Details on computing the CutX map weight function}
\label{sec:cutxweight}

\begin{figure}[ht!]
  \begin{center}
      \includegraphics[height=5cm]{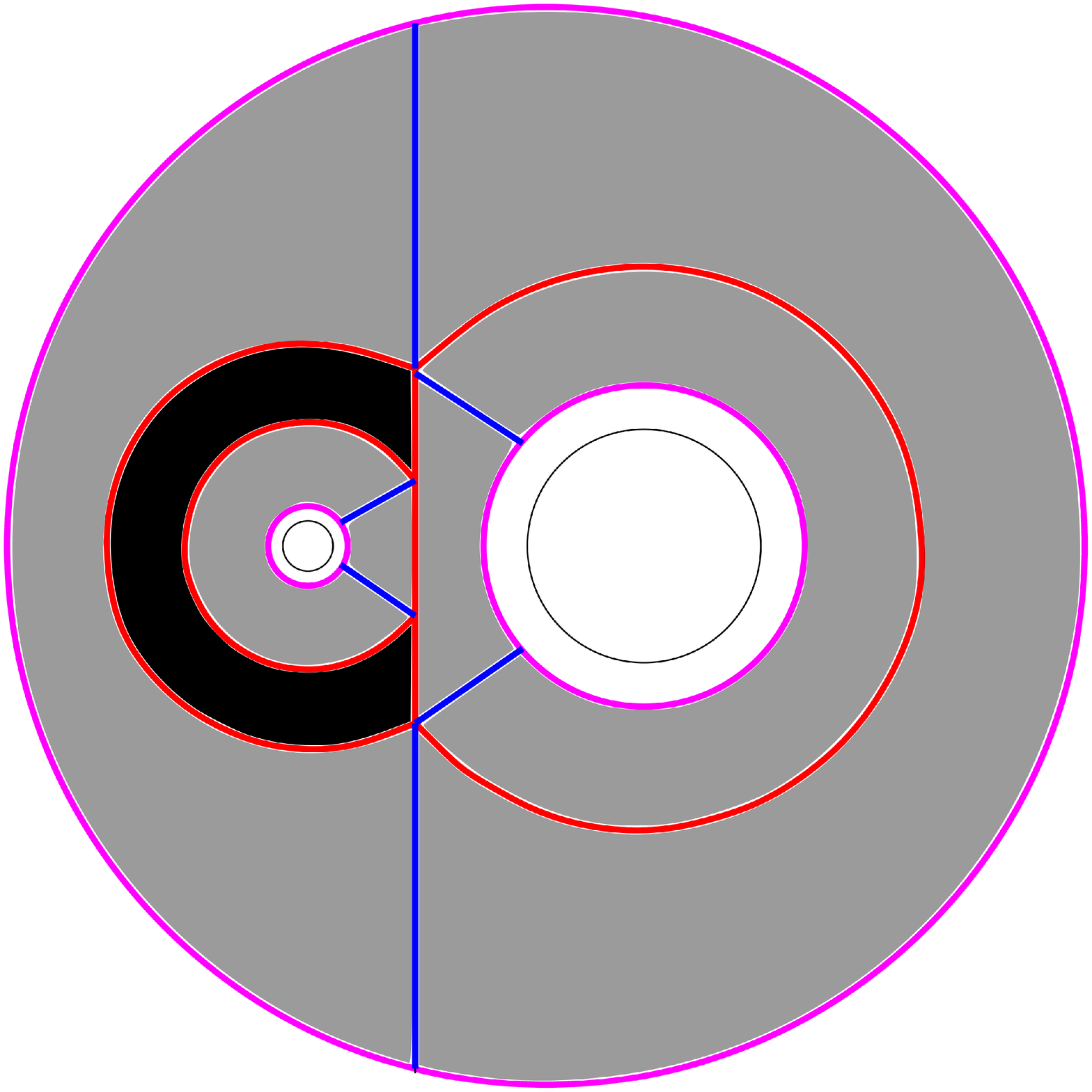}
  \end{center}
  \caption{Schematic diagram indicating
    the way the  \map{CutX} weight function
    is defined. The value of the weight function
    $\rho(x^i)$ is one on the red curves and
    in the black region enclosed by red curves,
    it vanishes in the white regions (inside the
    two inner magenta curves and outside the
    outer  magenta curve) and it changes
    linearly from zero to one in the gray regions
    between the red and magenta curves.}
  \label{fig:TranslationXForCutSphereWeight}
\end{figure}

As discussed in Sec.~\ref{sec:cutx}, we use the map \map{CutX}
to control the location of the cutting plane in the last phase
of the merger, as the distance between the excision boundaries
and the cutting plane becomes small.  The value of the weight function
$\rho(x^i)$ in Eq.~(\ref{eq:CutX.x}) is zero in the spherical shells
describing the wave zone and in the shells around the excision
boundaries (see Fig.~\ref{fig:TranslationXForCutSphereWeight}).
The weight function is unity
on the cutting plane, and on the ``cut-sphere''
surfaces around either excision boundary. For the smaller excision
surface, we have two such cut-sphere surfaces, and the value of the
weight function is one within the region enclosed by these two
cut-spheres and the cutting plane.  The weight function transitions
linearly from zero to one (from the magenta curves to the red curves
in Fig.~\ref{fig:TranslationXForCutSphereWeight}).
In these transitional regions the value of $\rho(x^i)$ is
computed as described below.

\begin{figure}[ht!]
  \begin{center}
      \includegraphics[height=5cm]{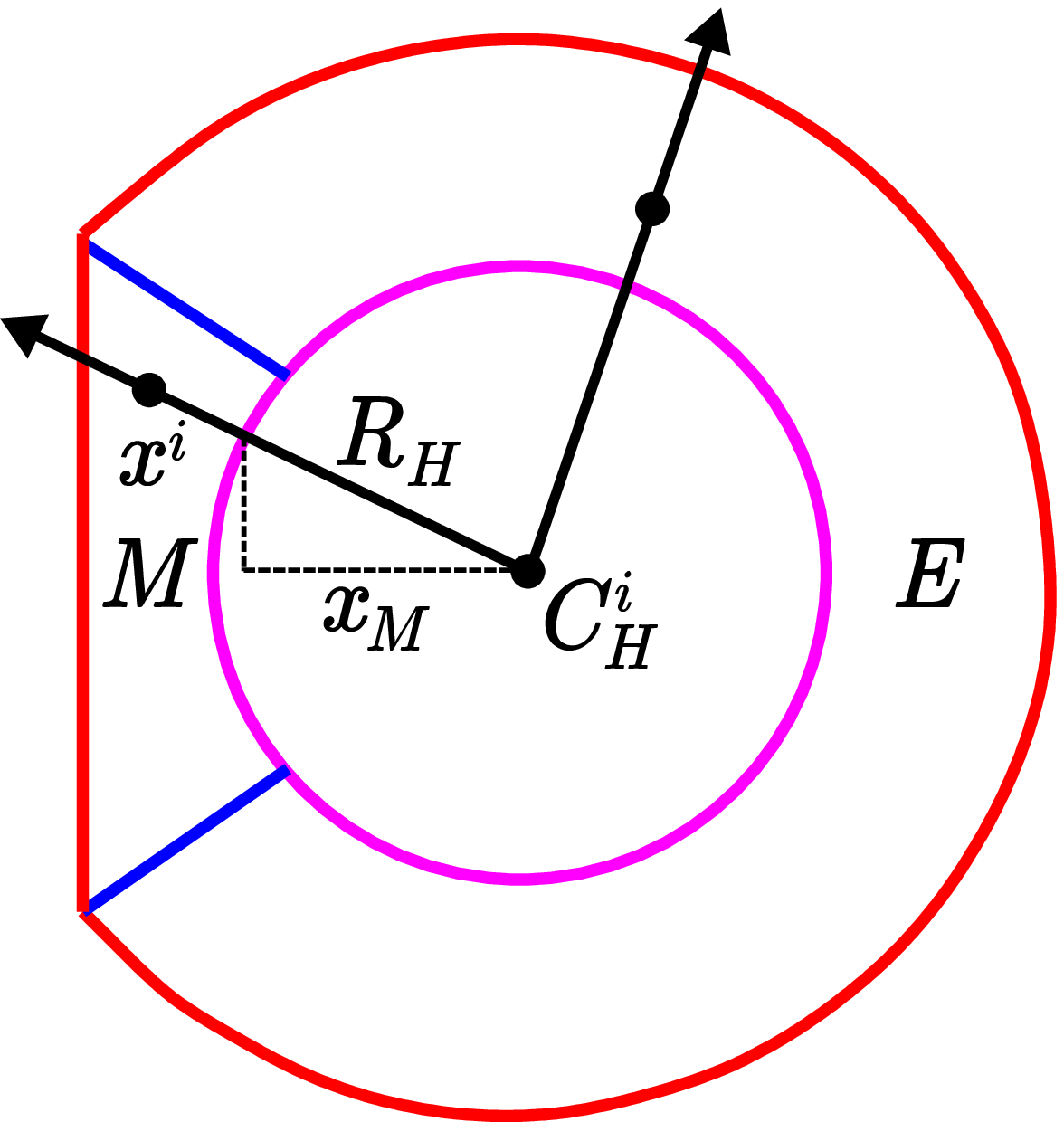}
  \end{center}
  \caption{
    Schematic diagram illustrating the algorithm
    used to compute the  \map{CutX} weight function
    inside the cut-sphere.
     The point $x^i$ represents an
    arbitrary point in the $M$ region.
    $R_H$ is the radius of the magenta sphere, and
    $x_M$ is the $x^0$-component of the
      point on the magenta sphere that intersects the ray pointing
      from $C_H^i$ toward $x^i$.
      The weight function $\rho(x^i)$
      is zero at the intersection
      of the ray with the magenta curve,
      it is unity at the intersection of the ray with
      the cutting plane,
      and changes linearly in between.
    Similarly, for a point in the $E$ region, between the magenta sphere
    and the spherical part of the
    red cut-sphere, 
     $\rho(x^i)$ changes linearly from zero
     (at the intersection of the ray with the inner, magenta sphere)
     to unity (at the intersection with the outer, red cut-sphere.)
}
  \label{fig:TranslationXForCutSphereWeightM}
\end{figure}

  Consider the region spanning the volume between
  the spherical shells around the excision boundary $H$
  (with $H=A,B$) and
  the cutting plane $x=x_C^0$, as shown
  in Fig.~\ref{fig:TranslationXForCutSphereWeightM}.
  (This is referred to in the code as the $M$ region.)
  In order to calculate the value
  of $\rho(x^i)$ in this region, we shoot a ray
from $C^i_H$
in the direction of $x^i$.
This ray intersects
  both the outer spherical boundary of the shells
  and the cutting plane.  The $x$-coordinate
  of the intersection with the shells will be
  \begin{equation}
    x_M = C^0_H + \frac{ (x^0-C^0_H) R_H }{ \sum_i\left(x^i-C_H^i\right)^2 },
  \end{equation}
  where $R_H$ is the outer radius of the spherical region around $C_H^i$.
   Then, for a point $x^i$ in the $M$ region $\rho$ is given by 
   \begin{equation}
    \rho(x^i) = \frac{x^0-x_M}{x^0_C-x_M}.
   \end{equation}

\begin{figure}[ht!]
  \begin{center}
      \includegraphics[height=5cm]{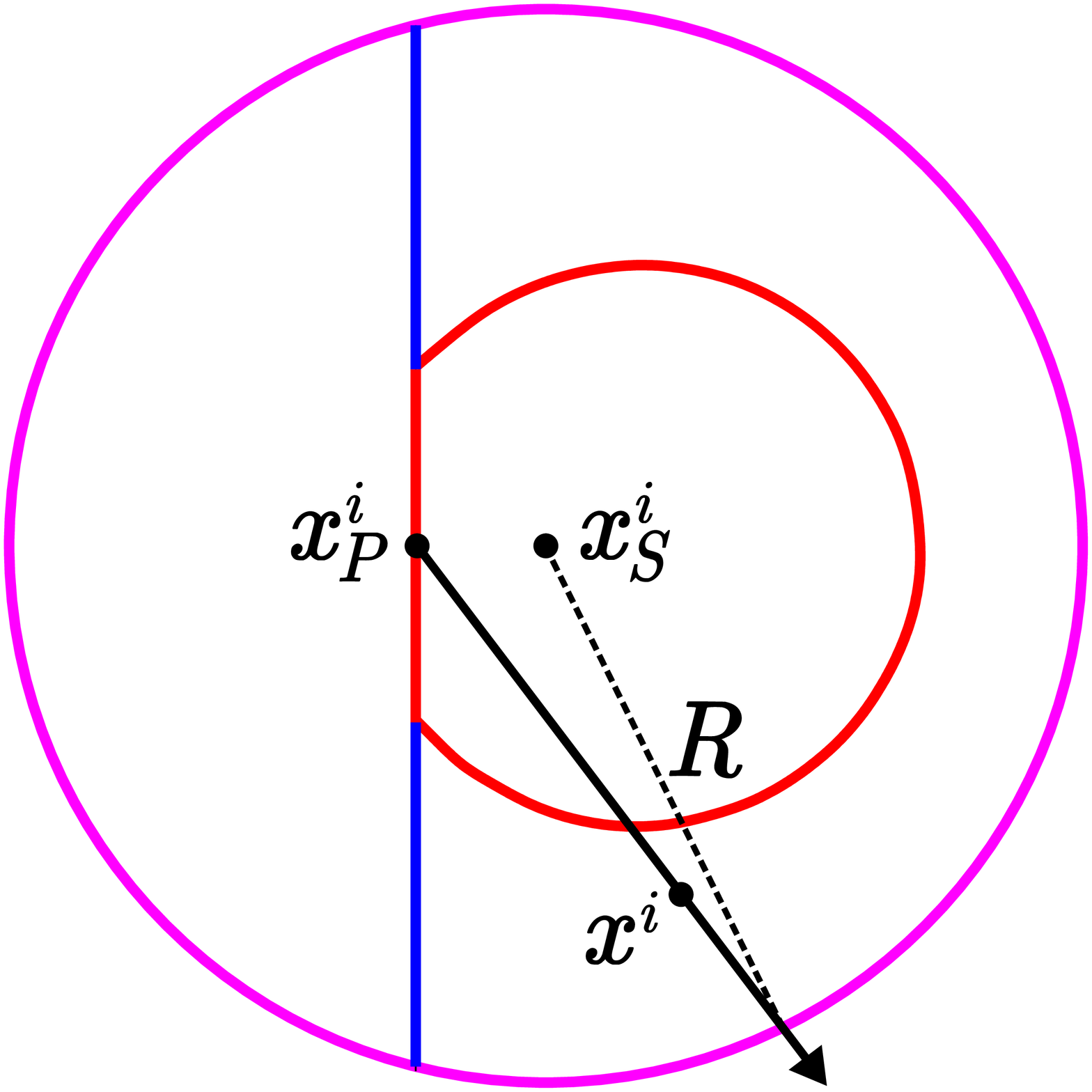}
  \end{center}
  \caption{
      Schematic diagram illustrating the algorithm used to
      compute the  \map{CutX} weight function in the region
      between the red cut-sphere and the outer, magenta sphere.
  }
  \label{fig:TranslationXForCutSphereWeightC}
\end{figure}

 All other transitional regions
   are delimited on both ends by spheres.
  Our computational domain has two zones of this type;
  one is depicted in Fig.~\ref{fig:TranslationXForCutSphereWeightM}
  (labeled as the $E$ region),
  while the other is seen in Fig.~\ref{fig:TranslationXForCutSphereWeightC}.
Let $x_P^i$ denote the ``center of projection,''
i.e., the
point
toward
which the unmapped radial grid lines of the given zone converge,
and let
$x_S^i$ denote the center of a sphere of radius $R$
(noting that the inner and outer spheres may have different centers).
  In Fig.~\ref{fig:TranslationXForCutSphereWeightM} the
  center of projection coincides with $C_H^i$.
  In Fig.~\ref{fig:TranslationXForCutSphereWeightC} we choose the
  center of projection to be on the cutting plane, at its intersection
  with the line segment connecting the centers of the two excision 
  surfaces, $x_C^i$.
To compute $\rho(x^i)$ in this type of region, we shoot a ray from $x_P^i$
in the direction of $x^i$. This ray intersects the two spheres that
delimit the region. We define
$r_0$ to be the distance between $x_P^i$ and the point of intersection
with the (magenta) sphere, for which $\rho=0$; similarly, we define $r_1$ to be the
distance between $x_P^i$ and the point of intersection with the (red) sphere,
for which $\rho=1$. Then, $\rho(x^i)$ is given by
\begin{equation}
 \rho(x^i) = \frac{r-r_0}{r_1-r_0},
\end{equation}
where $r = |x^i-x^i_P|$. To compute $r_0$ and $r_1$, we must solve
\begin{equation}
   \sum_i \left[ x_P^i + \eta (X^i - x_P^i) - x_S^i \right]^2 = R^2
\end{equation}
for the associated intersection point $X^i$. The parameter $\eta(x^i_P,x^i_S,X^i,R)$ is defined as
the positive solution of the quadratic system,
\begin{align}
\label{etaeq}
 \eta &= B + \sqrt{B^2-C},\\
\intertext{where}
 B &= \frac{\sum_i (X^i - x_P^i) (x_S^i-x_P^i)}{ \sum_i (X^i - x_P^i)^2},  \\
 C &= \frac{\sum_i  (x_S^i-x_P^i)^2 - R^2}{ \sum_i (X^i - x_P^i)^2}.
\end{align}

The input coordinates to \map{CutX} are the coordinates
$\frameB{x}^i$ that are obtained from the grid coordinates $x^i$ 
by the shape map, $\frameB{x}^i = \map{Shape}(x^i)$. 
As seen above, computation of the weight function at any point
requires knowledge of the region containing that point, which in turn
requires knowledge of its grid-frame coordinates $x^i$. Therefore,
to compute the weight function one must first compute
$x^i = \map{Shape}^{-1}(\frameB{x}^i)$.  When computing the
forward CutX map, this inverse 
is done only once.  However, the situation is more
complicated when computing the inverse CutX map, 
because the grid-frame coordinates depend upon the
inverse CutX map itself, which depends on the weight function.
This leads to an iterative algorithm
where we must call the inverse map function of \map{Shape} from each
iteration of the inverse map function of \map{CutX}. 
This could
easily become an efficiency bottleneck.  We mitigate this by keeping
\map{CutX} inactive for most of the run, only activating it when
it becomes crucial in order to avoid a singular map.


\section{Estimation of zero-crossing times}
\label{sec:estimatecrossing}

Several of the control systems described here are designed to ensure
that some measured quantity remains positive.  For example, in
Sec.~\ref{sec-AdaptiveSizeControl}, we demand that both the
characteristic speed $v$ at the excision boundary and the difference
$\Delta r$ between the horizon radius and the excision boundary radius
remain positive.  Similarly, in Sec.~\ref{sec:cutx} we demand that
each excision boundary does not cross the cutting plane.

Part of the algorithm for ensuring that some quantity $q$ remains
positive is estimating whether $q$ is in danger of becoming negative
in the near future, and if so, estimating the timescale $\tau$ on
which this will occur. In this Appendix we describe a method of
obtaining this estimate.

We assume the quantity $q$ is measured at a set of (not necessarily
equally-spaced) measurement times, and we remember the values of $q$ at
several (typically 4) previous measurement times.  At each measurement 
time $t_0$ we fit these remembered values to a line $q(t) = a
+ b(t-t_0)$.  The fit gives us not only the slope $b$ and the
intercept $a$, but also error bars for these two quantities $\delta a$
and $\delta b$.  Assuming that the true $q(t)$ lies within the error
bars, the earliest time at which $q$ will cross zero is
$t=t_0+(-a+\delta a)/(b-\delta b)$, and the latest time is
$t=t_0+(-a-\delta a)/(b+\delta b)$.  If both of these times are finite
and in the future,
then we regard $q$ as being in danger of crossing zero, and we
estimate the timescale on which this will happen as $\tau =
-a/b$.

\ack
We would like to thank Harald Pfeiffer for useful
discussions.  We would like to thank Abdul Mrou\'e 
for performing simulations that led to the development and improvement
of some of the mappings presented in this paper.  We gratefully
acknowledge support from the Sherman Fairchild Foundation; from NSF
grants PHY-0969111 and PHY-1005426 at Cornell, and from NSF grants
PHY-1068881 and PHY-1005655 at Caltech.  Simulations used in this work
were computed with the SpEC code~\cite{SpECwebsite}.  Computations
were performed on the Zwicky cluster at Caltech, which is supported by
the Sherman Fairchild Foundation and by NSF award PHY-0960291; on the
NSF XSEDE network under grant TG-PHY990007N; and on the GPC
supercomputer at the SciNet HPC Consortium~\cite{scinet}. SciNet is
funded by: the Canada Foundation for Innovation under the auspices of
Compute Canada; the Government of Ontario; Ontario Research
Fund--Research Excellence; and the University of Toronto.


\section*{References}  
\bibliographystyle{unsrt}
\bibliography{References/References}

\end{document}